\documentclass[twocolumn,aps,superscriptaddress,showkeys,showpacs]{revtex4-1}
\usepackage{graphicx}% Include figure files
\usepackage{dcolumn}% Align table columns on decimal point
\usepackage{bm}% bold math

\begin{document}

\preprint{CPB review}

\title{Crystal Chemistry and Structural Design of Iron-Based Superconductors}
\thanks{Project supported by the National Natural Science Foundation of China
(Grant Nos. 90922002 and 11190023) and the Fundamental Research Funds for the Central Universities of China (Grant No. 2013FZA3003)}%

\author{Hao Jiang}
\affiliation{Department of Physics, Zhejiang University, Hangzhou
310027, China}
\author{Yun-Lei Sun}
\affiliation{Department of Physics, Zhejiang University, Hangzhou
310027, China}
\author{Zhu-An Xu}
\affiliation{Department of Physics, Zhejiang University, Hangzhou
310027, China}
\author{Guang-Han Cao}
\email[Corresponding author. Email: ] {ghcao@zju.edi.cn}

\affiliation{Department of Physics, Zhejiang University, Hangzhou
310027, China}

%\date{\today}

\begin{abstract}
The second class of high-temperature superconductors (HTSCs),
iron-based pnictides and chalcogenides, necessarily contain
Fe$_2$$X_2$ ("$X$" refers to a pnictogen or a chalcogen element)
layers, just like the first class of HTSCs which possess the
essential CuO$_2$ sheets. So far, dozens of iron-based HTSCs,
classified into nine groups, have been discovered. In this article,
the crystal-chemistry aspects of the known iron-based
superconductors are reviewed and summarized by employing "hard and
soft acids and bases (HSAB)" concept. Based on these understandings,
we propose an alternative route to exploring new iron-based
superconductors via rational structural design.
\end{abstract}


\maketitle
\section{Introduction}
%short history of SC materials discovered
Since the discovery of superconductivity in the element mercury over
a hundred years ago,\cite{onnes} Numerous superconductors have been
found continually in a diversity of materials. These superconductors
can be classified into five groups, i.e., 1) elements, 2) alloys or
intermetallics, 3) inorganic compounds, 4) organic compounds, and 5)
polymers. Among them, the inorganic superconductors have been
studied most extensively and intensively in recent decades,
primarily because of the discovery of high temperature
superconductivity in cuprates\cite{bednorz,chu} and more recently in
iron pnictides\cite{hosono,cxh}.

While the nature of superconductivity was elucidated successfully
over half a century ago in terms of Cooper pairing mediated by
electron-phonon interactions,\cite{bcs} the theory by itself
supplies only limited guidance on exploring new superconductors, let
alone for the exotic superconductors beyond the electron-phonon
mechanism. Even if a material is theoretically designed and
calculated to be a desirable superconductors (e.g., with higher
superconducting transition temperature $T_c$), the proposed material
must be thermodynamically or at least kinetically stable, such that
it could be synthesized. The latter issue seems to be more crucial
and challenging. Under this circumstance, the knowledge of crystal
chemistry of a certain type of material is important and, it could
be helpful to look for new superconductors. As a matter of fact, in
the progress of finding new cuprate superconductors, the crystal
chemistry has played an indispensable role.\cite{raveau}

%discovery of iron-based SC
Iron-based superconductivity was first discovered in LaFePO in 2006
by H. Hosono and co-workers.\cite{hosono2006} The superconducting
transition temperature $T_c$ was only 3.2 K, which did not draw
instant attentions from the community of superconductivity. Two
years later, the same group reported superconductivity at 26 K in
LaFeAsO$_{1-x}$F$_x$.\cite{hosono} This breakthrough immediately
aroused great research interests. X. H. Chen et al.\cite{cxh}
successfully pushed the $T_c$ value beyond the McMillan limit by the
substitution of Sm for La, marking the birth of the second HTSCs.
The $T_c$ record of $\sim$55 K was also created at the
period.\cite{zzx,wc} In the following days of the year 2008, other
three types of Fe-based superconductors (FeSCs) were found one after
another.\cite{rotter,jcq,wumk} So far, the iron-based HTSC family
have included several dozens members. Although there have been a lot
of reviews on the subject of iron-based
superconductivity,\cite{rev-johnston,rev-stewart,rev-hosono,rev-zzx,rev-whh,rev-wilson,rev-greene,rev-chucw,rev-mazin,rev-ganguli}
a complete classification and description for the crystal structures
of FeSCs is still lacking. In this article, we present a
comprehensive review on the crystal chemistry of FeSCs, aiming to
explore new FeSCs in a rational way. We will not concentrate on the
structural details, and the structural phase transitions are simply
not touched. This related information can be found in Ref.~\cite{chem-johrendt}.

\section{Crystal chemistry}

\subsection{Chemical composition}

\begin{figure}
\includegraphics[width=7cm]{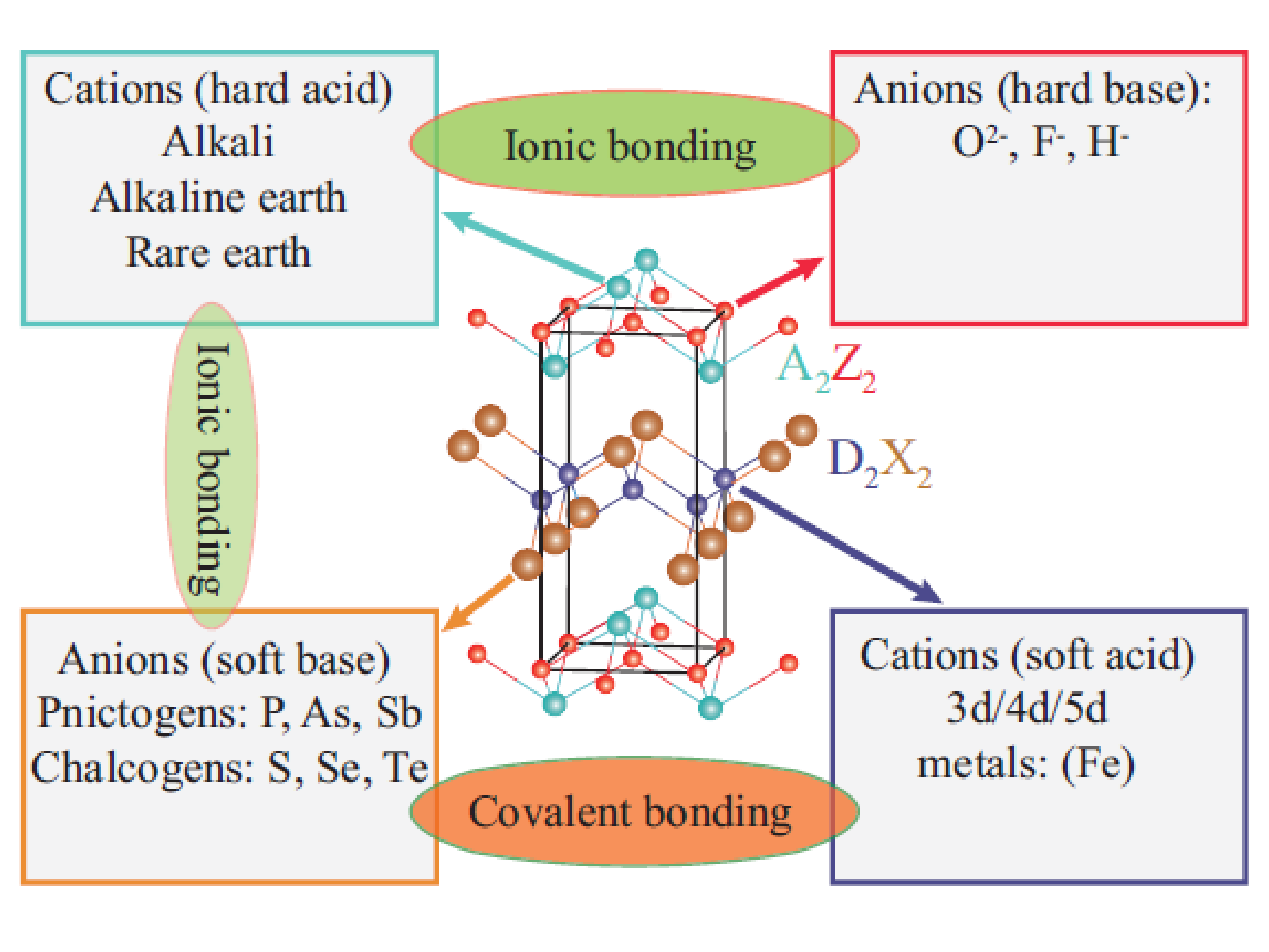}
\caption {\label{fig1}Characteristic of the constituent elements and
the chemical bonding in 1111-type $ADXZ$ iron-based
superconductors.}
\end{figure}

Crystal chemistry seeks for the principles describing
composition-structure-property relations in crystalline materials.
Let us first discuss features of the chemical composition of FeSCs.
Taken LaFeAsO$_{1-x}$F$_x$ (so-called "1111" material because it is
a quaternary equiatomic compound) for example, the constituent
elements (sequencing in the electronegativity order according to the
standard nomenclature of inorganic compounds) La$^{3+}$, Fe$^{2+}$,
As$^{3-}$ and O$^{-}$ belong to different groups that are called
hard acid, soft acid, soft base and hard base, respectively, in the
concept of "Hard and Soft Acids and Bases" (HSAB)\cite{pearson}. If
we employ a notation $ADXZ$ for the 1111 compounds, the
characteristic of the constituent elements is well defined: $A$
represents a "hard" (meaning non-polarizable) cation with the
smallest electronegativity; $D$ denotes a "soft" (meaning
polarizable) cation (it is normally a $d$-block transition metal,
but here it is virtually Fe for FeSCs); $X$ stands for a "soft"
anion (e.g., a pnictogen or a chalcogen); $Z$ is a "hard" anion with
the largest electronegativity. According to the HSAB
rule,\cite{pearson} $A$ tends to bond with $Z$ (an ionic bonding),
and $D$ combines with $X$ (covalent bonding dominated), as depicted
in Fig.~\ref{fig1}. The resultant two block layers $A_2$$Z_2$ and
$D_2$$X_2$ are linked by $A-X$ ionic bonding. As for other FeSCs,
the HSAB concept basically holds. So, we employ the
$A_{a}D_{d}X_{x}Z_{z}$ (simplified as $adxz$) notation throughout
this article.

The combination of $ADXZ$ leads to many members in the 1111 family.
A review paper published in 2008 lists over 150 $ADXZ$
individuals.\cite{pottgen} More 1111 compounds have been synthesized
since 2008. However, the number of the iron-based compounds reported
so far is only about 30. Importantly, the element-selective feature
at the four crystallographic sites allows various kinds of
successful chemical doping for inducing superconductivity. We will
have more discussions on these issues in the following sections.

\subsection{Fe$_2$$X_2$ layers represented by $\beta$-FeSe}

\begin{figure}
\includegraphics[width=7cm]{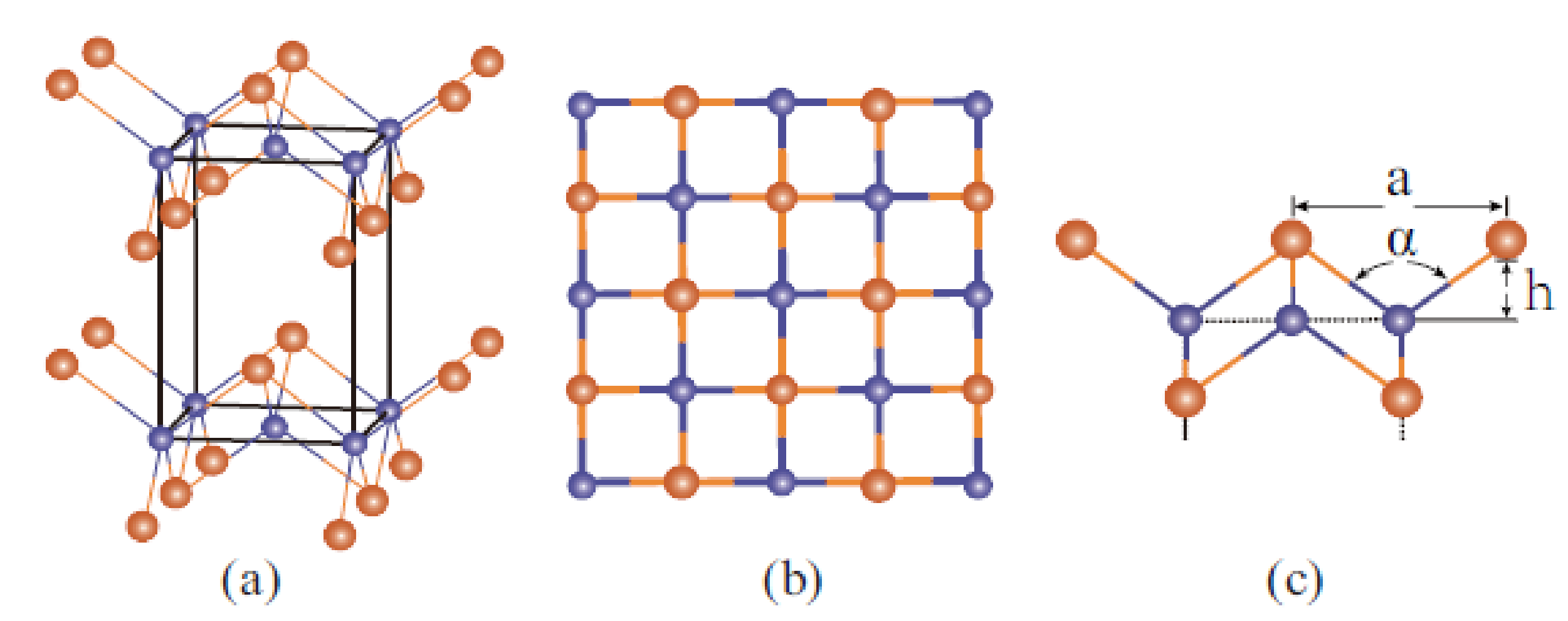}
\caption{\label{fig2}Crystal structure of $\beta$-FeSe which
consists of infinite Fe$_2$Se$_2$ layers. The structural parameters
that are considered to be crucial to the superconducting transition
temperature are marked in (c).}
\end{figure}

Just like cuprate superconductors that possess the essential CuO$_2$
sheets, all the known FeSCs necessarily contain Fe$_2$$X_2$ ("$X$"
refers to a pnictogen or a chalcogen element) layers. The crystal
structure of the Fe$_2$$X_2$ layers can be represented by
$\beta$-FeSe [so-called "11" phase, see Fig.~\ref{fig2}(a)] which
itself is a FeSC ($T_c$=8 K at ambient pressure).\cite{wumk} (There
was a confusion on the nomenclature in the literatures. FeSe with
nearly 1:1 stoichiometry actually crystallizes in two polymorphs.
One has a hexagonal NiAs-type structure, which is more stable than
the tetragonal phase. Normally the former is called $\alpha$-FeSe,
and the latter is called $\beta$-FeSe.\cite{FeSe-PD}) The
$\beta$-FeSe crystallizes in a layered anti-PbO-type structure with
space group $P4/nmm$. The Fe$_2$Se$_2$ monolayer consists of Fe$_2$
(two iron atoms in a unit cell) square-net sandwiched by two Se
monolayers. In the language of crystal chemistry, the Fe atoms are
coordinated by four Se atoms, and the resulted FeSe$_{4}$ tetrahedra
are edge-shared. In fact, the geometric configuration resembles that
of Li$_2$O (with anti-CaF$_2$-type structure). That is the reason
why literatures often refer the Fe$_2$$X_2$ layers as to
anti-fluorite-type structure.

Then, what $X$ can form stable Fe$_2$$X_2$ layers that incorporate
with other structural blocks to have a high-temperature FeSC?
Unfortunately, such $X$ is limited to Se and As so far, although
many efforts were made to explore possible superconductivity in an
iron compound with other $X$ such as Sb.\cite{X1,X2,X3}

Since the Fe$_2$$X_2$ layers are responsible for superconductivity,
the structural details of the Fe$X_{4}$ tetrahedra were considered
to be a determinant factor controlling the $T_c$. Empirically, the
relevant structural parameters are (1) the $a$
axis,\cite{renza,eisaki} (2) the bond angle of $X-$Fe$-X$
($\alpha$),\cite{zhaoj,lee} and (3) the height of $X$ with respect
to the Fe planes ($h$),\cite{mizuguchi} respectively. The latter two
parameters have been widely cited because the maximum $T_c$ in FeSCs
seems to fall at $\alpha$=109.5$^{\circ}$ (corresponding to a right
Fe$X_{4}$ tetrahedron) and $h$=1.38 {\AA}. Theoretical calculations
based on spin-fluctuation mechanism were able to elucidate the
variations of $T_c$ with increasing $h$.\cite{kuroki} Nevertheless,
overall $T_c$ does not obey a unique relation to any of these
parameters.\cite{rev-johnston} Perhaps it is simply not suitable to
consider the $T_c$ values of \emph{all} the FeSCs, irrespective of
doping level, doping site, uniaxial pressures, or isotropic
hydrostatic pressures. Besides, the two parameters $\alpha$ and $h$
are not irrelevant. Therefore, the empirical relations have obvious
flaws, which should be amended further.

Owing to the charge balance for satisfying Fe$^{2+}$, the 11-type
(tetragonal polymorph) iron-containing compound must be a
chalcogenide. So, there are only three members: FeS, FeSe and FeTe
in the 11 family. These simple binary compounds were reported as
early as in the 1930s.\cite{FeS,FeSe,FeTe} Among them, FeTe is the
most stable phase, and the crystals can be grown by a melting
method.\cite{FeTe-crystal} The $\beta$-FeSe polymorph is not so
stable. It was reported that annealing at lower temperatures was
absolutely necessary to obtain a single phase.\cite{rev-takano} The
anti-PbO-type FeS is simply metastable, and it can be synthesized
only through a soft chemical process in an aqueous
solution.\cite{FeS1995} The change in stability seems to be related
to the criteria/rule of ionic radius ratio.\cite{Pauling} The ionic
radius ratio between Fe$^{2+}$ and S$^{2-}$ does not satisfy the
tetrahedral coordination well. If the average size of the $X$-site
anions increases, the $\beta$-phase becomes stabilized. This is
manifested by the fact that it is easy to synthesize the solid
solutions of Fe(Te,Se)\cite{11-fmh,FeSeTe} and Fe(Te,S)\cite{FeTeS}.
Interestingly, $T_c$ can be enhanced up to about 14 K in both
systems.

Apparently the 11-type system is the simplest one among all the
FeSCs, however, here we emphasize that the real crystalline status
is much more complicated than one expected, primarily because some
iron atoms (in the form of Fe$^{2+}$) occupy the interstitial site
within the Van de Waals gap ($X$ bilayers). Most importantly, such
occupation suppresses superconductivity severely. It was reported
that only 3\% of the excess interstitial Fe completely destroyed the
superconductivity in Fe$_{1.03}$Se.\cite{11-cava}

\subsection{Other basic FeSC systems}

\begin{figure*}
\includegraphics[width=14cm]{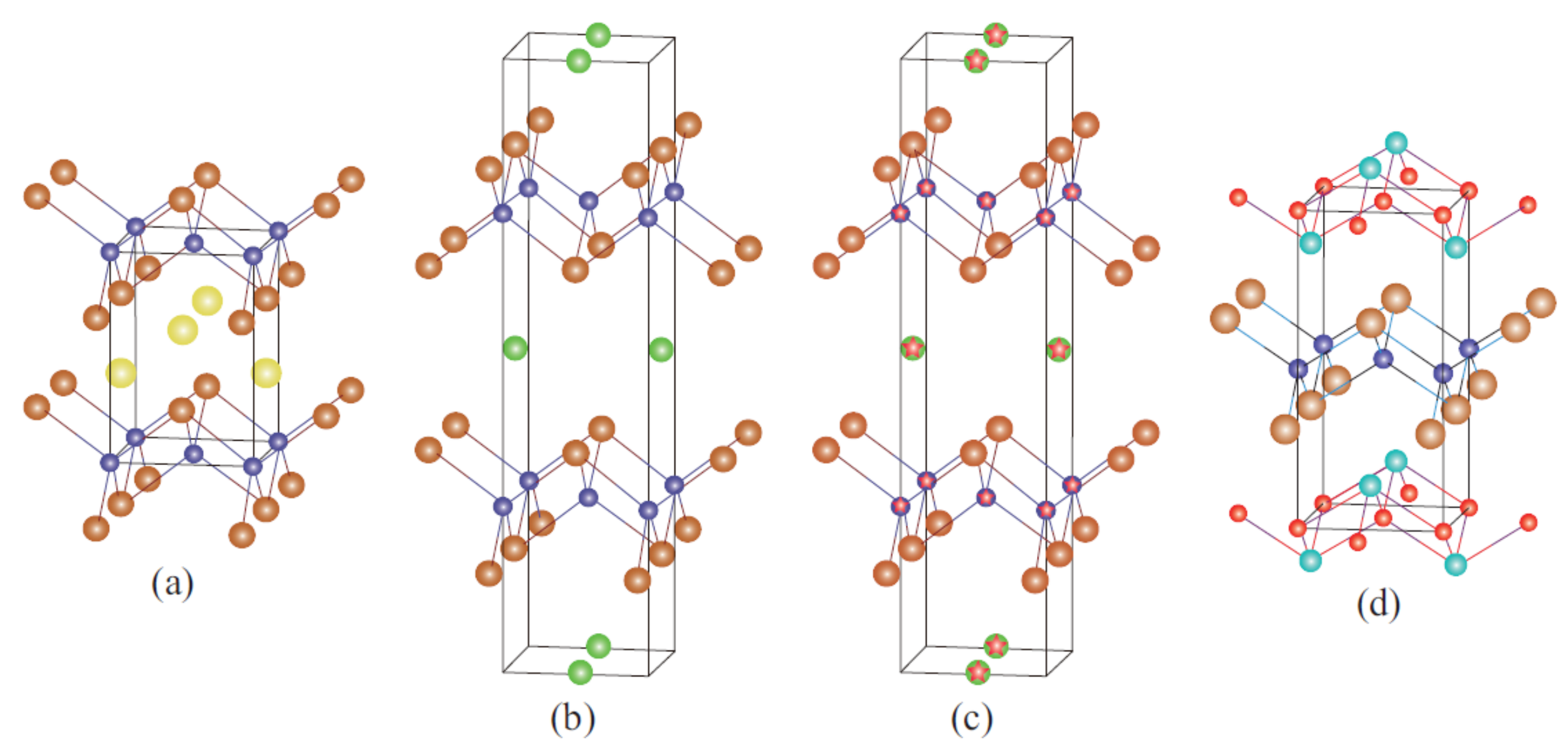}
\caption {\label{fig3}Basic crystal structures of iron-based
superconductors derived from the 11-type structure shown in Fig. 2.
(a) 111-type $ADX$; (b) 122-type $AD_{2}X_{2}$; (c) 122*-type
$A_{1-\delta}D_{2-\delta'}X_{2}$; (d) 1111-type $ADXZ$. See the text
for the notations.}
\end{figure*}

By inserting simple structural unit in between the Fe$_2$$X_2$
layers described above, some basic crystal structures can be derived
as shown in Fig.~\ref{fig3}. They are classified into the four
groups: namely, "111", "122", "122*"\cite{rev-stewart} and "1111"
systems, according to the above $adxz$ notation, which is consistent
with the common usage. The inserted units are a bilayer alkali and
monolayer (or partially occupied) alkaline earth for the 111 and 122
(122*) structure, respectively. As for the 1111 structure, the
$R_2$O$_2$ ($R$=rare earth) and $A_{E2}$F$_2$ ($A_{E}$=alkaline
earth) "block layers" are intergrown into the Fe$_2$$X_2$ layers.
Table~\ref{tab:table1} lists some representative FeSCs in the
category of the five basic structures.

\begin{table}
\caption{\label{tab:table1}Iron-based superconductors with five
basic structures (11, 111, 122, 122* and 1111). A general formula
$A_{a}D_{d}X_{x}Z_{z}$ (see the text) is used to distinguish the
type of elements that occupies/substitutes a certain
crystallographic site. Representative chemical doping strategies are
included. "HP" denotes the $T_c$ measured under high pressures.}

\begin{ruledtabular}
\begin{tabular}{lcr}
Structure/system&Chemical formula& $T_c$ (K) Refs.\\
\hline
11  & FeSe & 8\cite{wumk}\\
$DX$ & Fe(Se,Te) & 14\cite{11-fmh,FeSeTe}\\
    & FeSe & 37 (HP)\cite{11-HP1,11-HP2}\\
\hline
111 & LiFeAs & 18\cite{jcq}\\
$ADX$ & LiFeP & 6\cite{LiFeP-jcq}\\
\hline
 & Ba$_{1-x}$K$_{x}$Fe$_2$As$_2$ & 38\cite{rotter}\\
122& KFe$_2$As$_2$ & 3.8\cite{K122}\\
$AD_2$X$_2$  & BaFe$_2$As$_2$ & 29 (HP)\cite{122-hp}\\
 & BaFe$_{2-x}$Co$_{x}$As$_2$ & 25\cite{122Co}\\
 & BaFe$_2$As$_{2-x}$P$_{x}$ & 30\cite{js}\\
\hline
122* & K$_{1-x}$Fe$_{2-y}$Se$_2$ & 32\cite{cxl}\\
$A_{1-\delta}D_2$X$_2$  & (K,Tl)$_{1-x}$Fe$_{2-y}$Se$_2$ & 31\cite{122*-fmh}\\
 & (K,Tl)$_{1-x}$Fe$_{2-y}$Se$_2$ & 48 (HP)\footnotemark[1]\cite{sll}\\
\hline
 & LaFeAsO$_{1-x}$F$_{x}$ & 26\cite{hosono}\\
 & LaFeAsO$_{1-x}$F$_{x}$ & 41 (HP)\cite{1111-hp}\\
1111 & La$_{1-x}$Sr$_{x}$FeAsO & 25\cite{1111-whh}\\
$ADXZ$ & $R$FeAsO$_{1-x}$ & 40-55\cite{renza}\\
 & SmFeAsO$_{1-x}$F$_{x}$ & 55\cite{zzx}\\
 & Gd$_{0.8}$Th$_{0.2}$FeAsO & 56\cite{wc}\\
 & LaFe$_{1-x}$Co$_{x}$AsO & 14\cite{1111Co1,1111Co2}\\
 & LaFeAs$_{1-x}$P$_{x}$O & 10.5\cite{1111P}\\

\end{tabular}
\end{ruledtabular}
\footnotetext[1]{Another unknown superconducting phase at higher
pressures.}
\end{table}

The 111 phase crystallizes in an anti-PbFCl structure. Apparently it
differs much from the 11 phase, however, one may convert 11 to 111
continually by inserting alkali elements (Li or Na) into the
$X_5$-pyramid interstitial site. So, the space group is identical to
that of $\beta$-FeSe. Members of the $ADX$ family (with $D$=Fe) are
relatively limited due to the confinements of charge balance and
lattice match. Table~\ref{tab:table2} summarizes most of the
111-type iron compounds that have been synthesized.

\begin{table}[b]
\caption{\label{tab:table2}Iron-containing 111-type compounds.}
\begin{ruledtabular}
\begin{tabular}{ccccc}
$A$/$X$\footnotemark[1]&P&\text{As}&\text{Si}&\text{Ge}\\
\hline
Li&LiFeP\cite{LiFeP-jcq}&LiFeAs\cite{111str}&$---$&$---$\\
Na&$---$&NaFeAs\cite{NaFeAs-wnl}& $---$ & $---$ \\
Mg&$---$&$---$ &$---$&MgFeGe\cite{MgFeGe} \\
$R$&$---$&$---$ & $R$FeSi\footnotemark[2]&$---$\\
\end{tabular}
\end{ruledtabular}
\footnotetext[1]{CuFeSb also belongs to the 111 family. It is a rare
iron antimide without obvious Fe deficiency. Ferromagnetism with a
Curie temperature of 375 K was recently reported.\cite{CuFeSb-mzq}}
\footnotetext[2]{$R$=Y, Ce, La, etc.}
\end{table}

It is noted that, among the compounds listed in
Table~\ref{tab:table2}, only the pnictides show superconductivity so
far. LiFeP and LiFeAs become superconducting at 6 K and 18 K,
respectively.\cite{jcq,LiFeP-jcq} In fact, all the iron phosphides
known have relatively low $T_c$. Nevertheless, it is not usual that
the undoped LiFeAs shows superconductivity, because most undoped
iron pnictides like NaFeAs exhibit antiferromagnetic
spin-density-wave transition.\cite{rev-johnston,rev-stewart} The
possible reason is that LiFeAs has a very small lattice
($a\sim$3.77{\AA}). This implies that the "internal pressure" is at
work, which leads to superconductivity by itself. Similar
explanation might be valid for the appearance of superconductivity
in $\beta$-FeSe. In comparison, the undoped NaFeAs shows complex
magnetic transitions and, it is not superconducting without
extrinsic doping. Superconductivity at 23 K appears when the sodium
is significantly deficient.\cite{NaFeAs-wnl} By the Co doping and/or
with applying high pressures, the $T_c$ can be increased up to 31 K,
the record to our knowledge in the 111 system.\cite{111-cxh,111-hp}

The absence of superconductivity in the 111-type iron silicides and
germanides\cite{MgFeGe-hosono} deserves further study. Although the
Ge$-$Fe$-$Ge bond angle ($\alpha$=103.55$^{\circ}$) is far less than
the ideal value, it is actually larger than that of LiFeAs
($\alpha$=102.79$^{\circ}$\cite{111-pitcher}). Therefore, some other
factors such as electron correlation effect may also crucial for the
appearance of superconductivity.

The 122 $AD_{2}X_{2}$ compounds have the ThCr$_2$Si$_2$-type
structure with $I4/mmm$ space group. The $A$-site cations are
coordinated by eight $X$ atoms, thus the $X-A-X$ triple-layer unit
is analogous to CsCl-type block. So, this geometric configuration
requires relatively large cations for $A$, as is true from
Table~\ref{tab:table3}. For the pnictides, $A$ can be a larger
alkali or alkaline earth elements. Eu is the exceptional rare earth
that can form 122 ferroarsenides because Eu$^{2+}$ is relatively
stable and has relatively large size. The most interesting point for
the Eu-containing 122 phases is that they may show anomalous
superconducting and magnetic properties, which were called Fe-based
ferromagnetic superconductors.\cite{renz,jiangs,jwh,cao} The
researches along this line are worthy and expectable.

%122 Table of 122
\begin{table}[b]
\caption{\label{tab:table3}%
Iron-containing 122 $A$Fe$_{2}X_{2}$ compounds. The related
references can be seen in a review article.\cite{122review} The
review lists about 600 members in the ThCr$_2$Si$_2$-type family.}
\begin{ruledtabular}
\begin{tabular}{ccccc}
$A$/$X$&P&\text{As}&\text{Si}&\text{Ge}\\
\hline
K&KFe$_2$P$_2$&KFe$_2$As$_2$&$---$&$---$\\
Rb&$---$&RbFe$_2$As$_2$&$---$&$---$\\
Cs&CsFe$_2$P$_2$&CsFe$_2$As$_2$&$---$&$---$\\
\hline
Ca&CaFe$_2$P$_2$&CaFe$_2$As$_2$&$---$&$---$\\
Sr&SrFe$_2$P$_2$&SrFe$_2$As$_2$&$---$&$---$\\
Ba&BaFe$_2$P$_2$&BaFe$_2$As$_2$&$---$&$---$\\
\hline
Eu&EuFe$_2$P$_2$&EuFe$_2$As$_2$&EuFe$_2$Si$_2$&EuFe$_2$Ge$_2$\\
\hline
$R$&$R$Fe$_2$P$_2$&$---$&$R$Fe$_2$Si$_2$&$R$Fe$_2$Ge$_2$\\
\end{tabular}
\end{ruledtabular}
\end{table}

ThCr$_2$Si$_2$-type structure is widely adopted. A review paper in
1996 listed nearly 600 compounds of the type.\cite{122review} For
the iron-based compounds, the number was much decreased. Even so,
like the 111-type material, only the pnictides show
superconductivity. The phosphides generally have very low $T_c$. For
the arsenides, the $T_c$ value depends on the doping level. In
Sr$_{1-x}$K$_{x}$Fe$_2$As$_2$ system, for example, the undoped
SrFe$_2$As$_2$ is not superconducting. With increasing the K
substitution, $T_c$ increases to a maximum value of 37 K at $x$=0.4,
and then $T_c$ decreases steadily down to 3.8 K for
$x$=1.\cite{K122} In an electron-doped
Ca$_{1-x}$La$_{x}$Fe$_2$As$_2$ system, superconducting transition up
to 49 K was observed.\cite{CaLa122}

The $AD_{2}X_{2}$ compounds can be classified into two categories,
depending on whether $X-X$ is bonding or not.\cite{122collapse} For
the compound with a small $c/a$ ratio ($\sim$2.8), the $X-X$ bonding
is significant and, one often calls it a collapsed phase. It is
interesting that no collapsed phases show high-temperature
superconductivity, irrespective of chemical doping or applying
pressures. This is probably related to the covalent $X-X$ bonding,
which greatly influences the electron correlations, the Fermi levels
as well as the Fermi surface topology. Generally the collapsed phase
has a relatively large $\alpha$ angle, which does not favor high
$T_c$ superconductivity. So, it is not surprising that the iron
silicides and germanides do not show superconductivity at an
elevated temperature.

There is another common structure type called
CaBe$_2$Ge$_2$\cite{CaBeGe}, whose stoichiometry is also 1:2:2. It
is also tetragonal, and contains anti-fluorite-like Be$_2$Ge$_2$
layers. The different structural feature is that there is a reverse
arrangements for Be and Ge atoms, which forms Ge$_2$Be$_2$ layers
that are linked with the "normal" Be$_2$Ge$_2$ block by Ca$^{2+}$
cations. It was recently discovered that SrPt$_2$As$_2$,
crystallized in the CaBe$_2$Ge$_2$-type structure, shows
superconductivity at 5.2 K.\cite{SrPtAs} Unfortunately it seems that
the 4$d$ and 5$d$ elements tend to adopt this kind of structure, and
no iron-based compounds with this structure has been reported so
far.

At the end of 2010, Guo et al.\cite{cxl} reported a 122-type
ferroselenide superconductor K$_{x}$Fe$_2$Se$_2$ with $T_c$=32 K.
While the average structure is the same as a normal ThCr$_2$Si$_2$,
the cation deficiency, especially at the Fe-site,\cite{122*-fmh}
makes the system special. So, it was called 122* phase by
Stewart.\cite{rev-stewart} There have been a lot of studies along
this line, primarily because it has not only interesting
superstructures but also complicated phase separations. For readers
who are interested in the details, we recommend a recent
review\cite{rev-whh2012} which describes the main developments and
the perspectives for this particular system.

The quaternary equiatomic compounds $ADXZ$ have a tetragonal
ZrCuSiAs-type structure ($P4/nmm$). As stated in the above section,
the combination of $ADXZ$ may produce many 1111-type
compounds.\cite{pottgen} Nevertheless, the iron-based 1111 compounds
are not so wealthy, as listed in Table~\ref{tab:table4}. Here we
note that most of the ferro-oxyarsenides were synthesized by Quebe
et al in 2000.\cite{1111-quebe} If these compounds had been well
studied in time, FeSCs would come out much earlier.

%1111 Table of 1111

\begin{table}[b]
\caption{\label{tab:table4}Iron-containing 1111 $ADXZ$ compounds.
Most of the related references can be seen in a review
article.\cite{pottgen} The review lists over 150 members in the
ZrCuSiAs-type family.}
\begin{ruledtabular}
\begin{tabular}{cccc}
$A$-$Z$/$X$&P&\text{As}&Si\\
\hline
$R$\footnotemark[1]-O&$R$FePO&$R$FeAsO&$---$\\
\hline
Np-O&$---$&NpFeAsO\cite{Np1111}&$---$\\
\hline
Ca-F&$---$&CaFeAsF&$---$\\
Sr-F&$---$&SrFeAsF&$---$\\
Ba-F&$---$&$---$&$---$\\
Eu-F&$---$&EuFeAsF&$---$\\
\hline
Ce-H&$---$&$---$&CeFeSiH\\
Ca-H&$---$&CaFeAsH\cite{1111H-hosono}&$---$\\
\end{tabular}
\end{ruledtabular}
\footnotetext[1]{$R$=La, Ce, Pr, Nd, Sm, Gd (synthesized at ambient
pressure); More compounds with $R$=Tb, Dy, Ho, etc, were obtained by
high-pressure synthesis technique.}
\end{table}

\subsection{Extended FeSC systems: intergrowth structures}

\subsubsection{Intergrowth with perovskite-like block layers}

\begin{figure*}
\includegraphics[width=14cm]{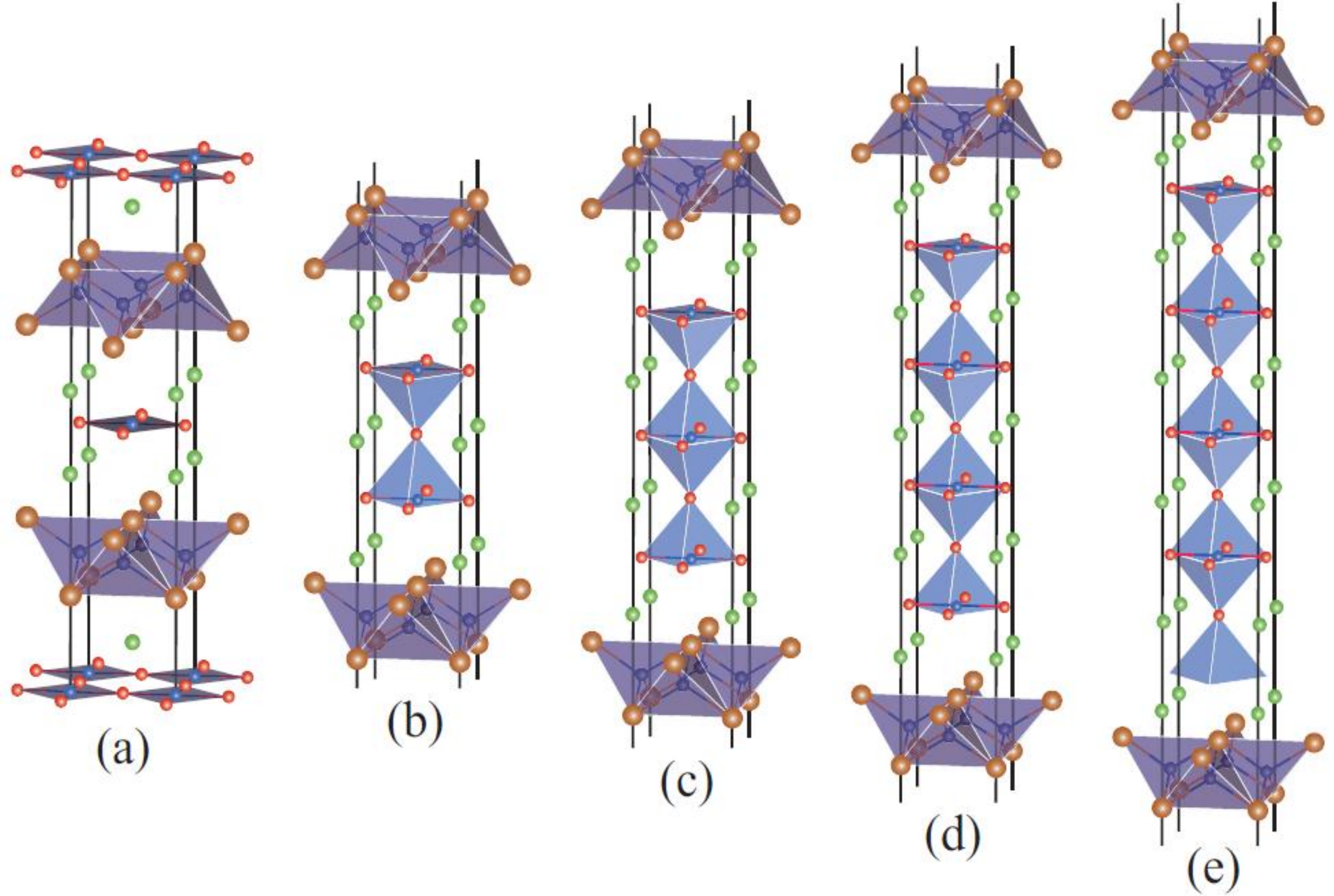}
\caption {\label{fig4}Iron-based superconductors with intergrowth
structures containing perovskite-like block layers. The general
formula is $A_{n+1}B_{n}D_{2}X_{2}Z_{3n-1}$. The structures
displayed from (a) to (e) correspond to $n$=1 to 5.}
\end{figure*}

The $a$ axis of the above FeSCs ranges from 3.77 {\AA} to 4.03
{\AA}, which makes $AB$O$_{3-\delta}$ perovskite-like layers
compatible to form an intergrowth structure. As a matter of fact,
similar intergrowth structures were reported
earlier.\cite{21222-zhu,32225-zhu,43228-zhu,32225-kishio,43228-hyett}
For example, the crystal structure of Sr$_2$$B$O$_2$Cu$_2$S$_2$
($B$=Zn and Co) is an intergrowth of 122-type SrCu$_2$S$_2$ and
oxygen-deficient perovskite-like Sr$B$O$_2$ infinite
layers.\cite{21222-zhu} Other intergrowth structures with thicker
perovskite layers were also developed over a decade
ago.\cite{32225-zhu,43228-zhu} The homologous series can be written
as $A_{n+1}B_{n}D_{2}X_{2}Z_{3n-1}$, whose crystal structures were
shown in Fig.~\ref{fig4}. The $n$=2 structure consists of
$A_{3}B_{2}Z_{5}$ block layers, and the $n$=3 phase has more thicker
of $A_{4}B_{3}Z_{8}$ building block.
Sr$_3$Fe$_2$Cu$_2$S$_2$O$_5$\cite{32225-zhu} and
Sr$_3$Sc$_2$Cu$_2$S$_2$O$_5$\cite{32225-kishio} belong to the $n$=2
member, and
Sr$_4$Mn$_3$Cu$_2$S$_2$O$_{8-\delta}$\cite{43228-zhu,43228-hyett} is
for the case of $n$=3.

The first reported compound with Fe$_2$As$_2$ and perovskite-like
layers was 32225 (it was called 32522 in some literatures)
ferroarsenide Sr$_3$Sc$_2$Fe$_2$As$_2$O$_5$,\cite{32225-whh} an
$n$=2 member. Although the material does not show spin-density-wave
anomaly, it was considered to host potential superconductivity.
Indeed, traces of superconductivity at about 20 K were observed in
the Ti-doped 32225 ferroarsenide.\cite{32225-chengf}. Recently, the
32225 ferropnictides Ca$_3$Al$_2$Fe$_2$$X_2$O$_{5-\delta}$ ($X$=As
and P), synthesized at high pressures, were reported to show
superconductivity at 30.2 K and 16.6 K,
respectively.\cite{32225-shirage} Other intergrowth structures with
thicker perovskite block layers ($n$=3,4,5) were also synthesized,
and they all exhibit
superconductivity.\cite{42238-ogino,thicker-ogino}

There was one special member of $n$=1: Sr$_2$CuO$_2$Fe$_2$As$_2$
which comprises of superconductively-active Fe$_2$As$_2$ layers and
CuO$_2$ sheets. Our first-principles calculations suggest a charge
transfer from the Fe$_2$As$_2$ layers to the CuO$_2$
sheets.\cite{master-thesis} So, it is expected to show successive
superconducting transitions in the two layers. Probably many people
tried to synthesize this hypothetical object. The effort led to some
related 21222 phase (with mixed occupations at $B$ and $D$
sites),\cite{21222-nath} but the target material has never been
obtained. Very recently, Sr$_2$CrFe$_2$As$_2$O$_2$ was successfully
synthesized, but unfortunately the material was not superconducting
down to 3 K.\cite{21222Cr}

\begin{figure}
\includegraphics[width=7cm]{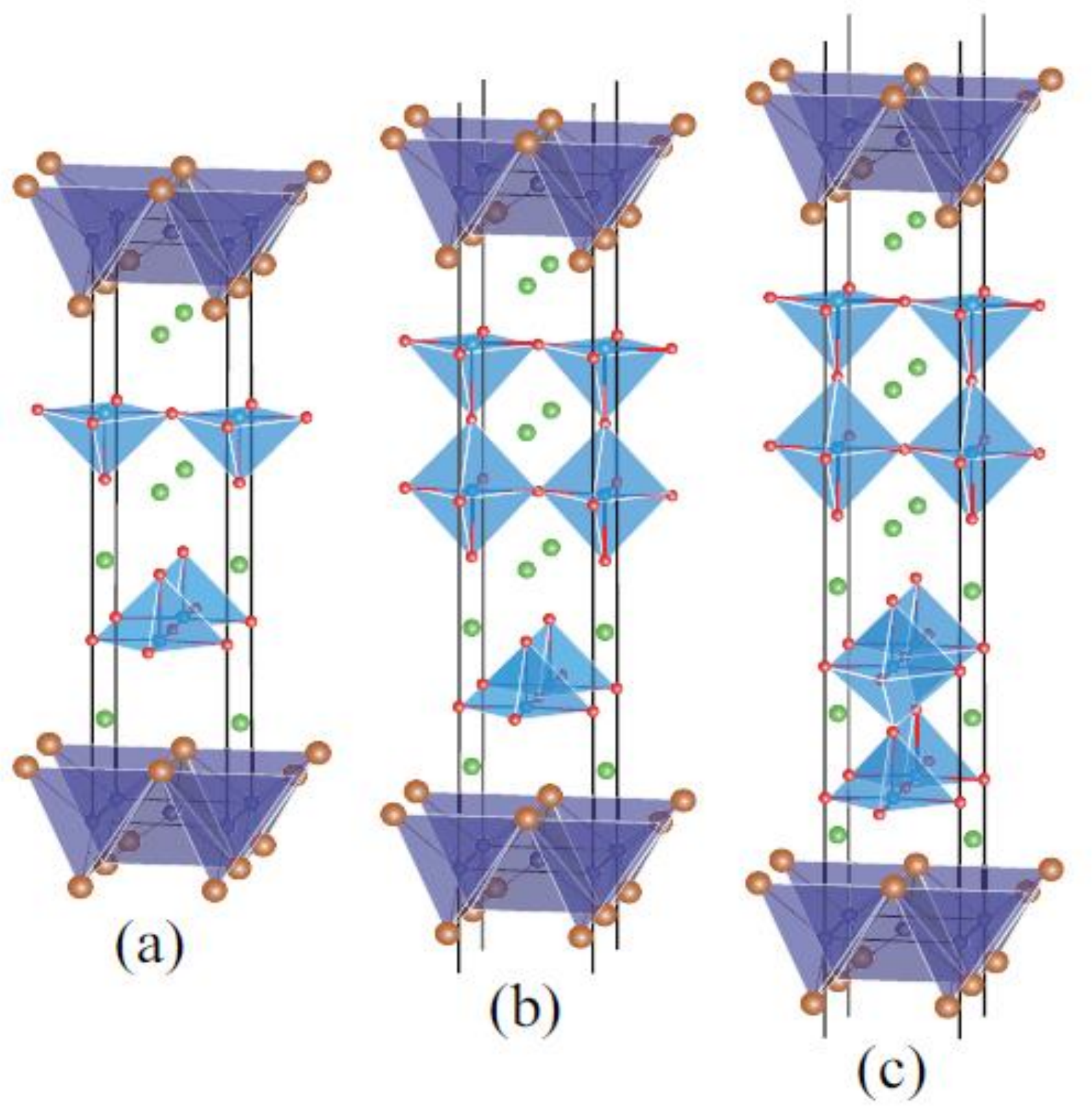}
\caption {\label{fig5}Crystal structures of a homologous iron-based
superconductors $A_{n+2}B_{n}D_{2}X_{2}Z_{3n}$ containing
Ruddlesden-Popper block layers. Three members with $n$=2, 3 and 4
have been synthesized.\cite{53229}}
\end{figure}

%21113
There is a perovskite derivative called Ruddlesden-Popper (RP)
series $A_{n+1}B_{n}Z_{3n+1}$, which contain a rock-salt layer in
addition to the perovskite layer.\cite{RP} Similarly, the
oxygen-deficient RP layers can also be intergrown with the 122-type
structure, forming another homologous series
$A_{n+2}B_{n}D_{2}X_{2}Z_{3n}$. The crystal structures for $n$=2, 3
and 4 were shown in Fig.~\ref{fig5}.

The pioneering example in this series was Sr$_2$GaCuSO$_3$ which
consists of Sr$_3$Ga$_2$O$_6$ RP block and SrCu$_2$S$_2$ structural
unit.\cite{21113-zhu} It was found that the 21113 (it was called
21311 or 42622 in some literatures) ferro-oxyphosphide
Sr$_2$ScFePO$_3$ shows superconductivity at 17 K, the highest $T_c$
in phosphides so far.\cite{21113P} For the arsenides,
Sr$_2$VFeAsO$_3$ exhibits superconductivity at 37.2 K without
extrinsic doping or applying pressures.\cite{21113-whh} The present
author and co-workers explained it in terms of electronic
self-doping due to a spontaneous charge transfer between the two
building blocks.\cite{21113-cao} In the cases of $B$=Sc and Cr, the
charge transfer is not obvious, thus they did not show
superconducting transitions.\cite{21113Sc, 21113Cr} Partial
substitution of Sc by Ti in Sr$_2$ScFeAsO$_3$ leads to
superconductivity up to 45 K.\cite{32225-chengf}

In this homologous series, $n$=3 and 4 members were also
successfully obtained by Ogino et al with $A$=Ca,
$B$=Al/Sc/Ti.\cite{53229} The $n$=4 phase was well separated and
characterized. Notably, the $a$-axis the synthesized compounds are
approximately 3.8 {\AA}, close to those of FeSe and LiFeAs. Besides,
the incorporation of Ti essentially induces excess electrons into
the Fe$_2$As$_2$ layers. Thus the series of new compounds exhibited
bulk superconductivity with $T_c$ up to 39 K.

\subsubsection{Intergrowth with other exotic block layers}

\begin{figure}
\includegraphics[width=7cm]{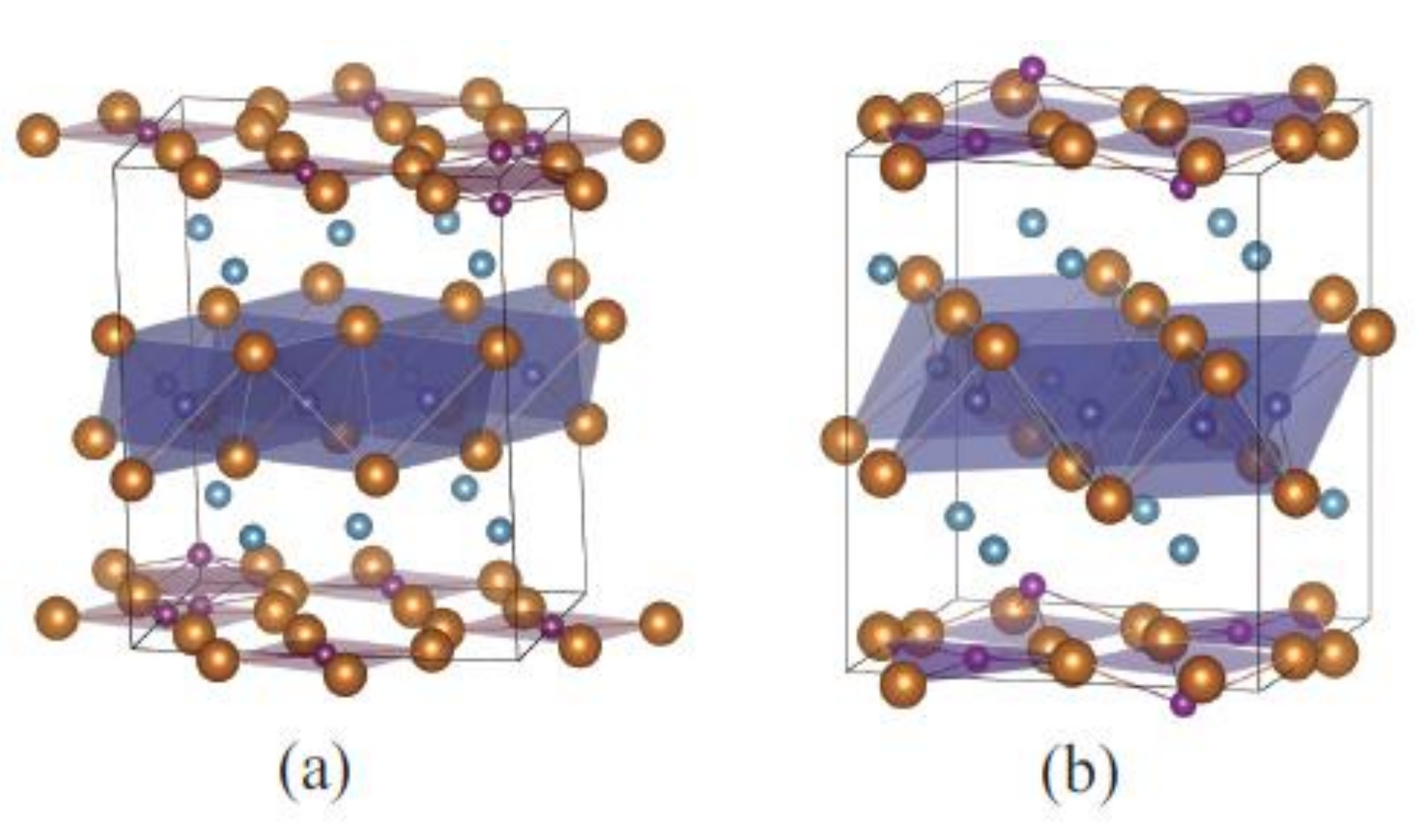}
\caption {\label{fig6}Crystal structures of iron-based
superconductors containing intermediate layers of Pt$_{3}$As$_8$ and
Pt$_{4}$As$_8$ (based on a $\sqrt{5}\times\sqrt{5}$ superlattice in
Fe$_2$As$_2$ layers). Shown in (a) is for
Ca$_{10}$Pt$_3$As$_8$(Fe$_2$As$_2$)$_5$, and (b) is for
Ca$_{10}$Pt$_4$As$_8$(Fe$_2$As$_2$)$_5$.}
\end{figure}

In addition to the perovskite-like layers, some "exotic" block
layers were also able to be inserted into the Fe$_2$$X_2$ layers. In
2011, two platinum-containing compounds
Ca$_{10}$Pt$_3$As$_8$(Fe$_2$As$_2$)$_5$ and
Ca$_{10}$Pt$_4$As$_8$(Fe$_2$As$_2$)$_5$, called 10-3-8 and 10-4-8
phases respectively, were reported.\cite{Pt-cava} The crystal
structures (Fig.~\ref{fig6}) are not so complicated as the formula
suggest. The Fe$_2$As$_2$ layers are virtually intact, which makes
it possible to become superconducting through a chemical doping. The
intermediate part is a little complex. The Pt atoms are
square-coordinated, and resultant PtAs$_4$ squares are linked by
sharing the vertex. The 2$\times$2 superlattice of Pt (i.e.,
Pt$_4$As$_8$) happens to match the $\sqrt{5}\times\sqrt{5}$
superlattice of Fe$_2$As$_2$, forming 10-4-8 structure. In the
intermediate layer of the 10-3-8 phase, one of the four Pt sites is
absent, leaving the Pt$_3$As$_8$ planes between Fe$_2$As$_2$ layers.

Through partial substitution of Pt for Fe in the Fe$_2$As$_2$
layers, superconductivity at 11 K and 26 K were observed in the
10-3-8 and 10-4-8 phases, respectively.\cite{Pt-cava} The difference
in $T_c$ for these similar compounds with similar As$-$Fe$-$As bond
angle challenges the empirical rule and, it was suggested that
interlayer coupling plays an important role in enhancing $T_c$ in
the FeSCs.

\begin{figure}
\includegraphics[width=5.5cm]{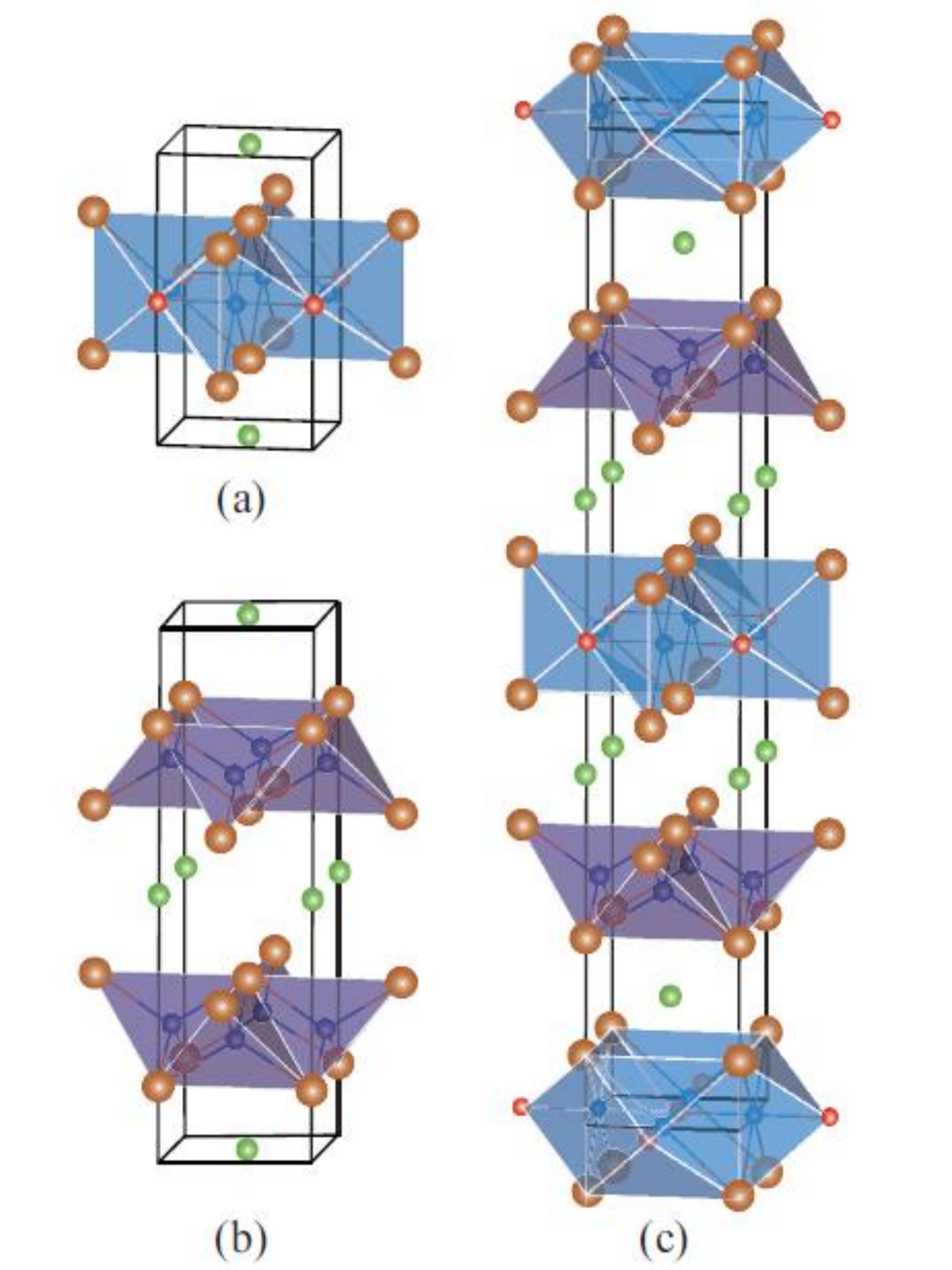}
\caption {\label{fig7}Intergrowth of BaTi$_{2}$As$_{2}$O (a) and
BaFe$_{2}$As$_{2}$ (b) produces a new iron-based superconductors
Ba$_{2}$Ti$_{2}$Fe$_{4}$As$_{4}$O (c). Note that the
Ti$_{2}$As$_{2}$O layers are also conducting, making the material
distinct from other FeSCs.}
\end{figure}

We noted that a class of titanium oxypnictides contains Ti$_{2}$O
square lattice,\cite{ozawa-review} which well matches Fe$_2$As$_2$
layers. The interesting issue of the titanium oxypnictides lies in a
density-wave (DW) anomaly that is related to the Ti$_{2}$O sheets.
In 2010, a relatively simple titanium oxypnictide
BaTi$_{2}$As$_{2}$O was prepared and characterized, which shows a DW
anomaly at 200 K.\cite{1221-cxh} We then considered to incorporate
the Ti$_{2}$O sheets into the Fe$_2$As$_2$ layers. After some
attempts, we finally succeeded in synthesizing the intergrowth
structure of BaTi$_{2}$As$_{2}$O and BaFe$_{2}$As$_{2}$,\cite{syl}
as shown in Fig.~\ref{fig7}. Bulk superconductivity at 21 K was
observed after the sample annealing without apparent doping. In
addition, a DW anomaly appears at $T_{DW}$=125 K, which is
remarkably lower than that of BaTi$_{2}$As$_{2}$O. By a
first-principles calculation, these phenomena were attributed to a
self doping due to the charge transfer from the Ti$_{2}$O sheets to
the Fe$_2$As$_2$ layers.\cite{jh} Further physical property
investigations are under way.

Table~\ref{tab:table5} summarizes the extended four types of FeSCs
with relatively thick intermediate layers. The optimal $T_c$ values
seem to be independent of the type of intermediate layers. Although
the empirical rule for $T_c$ is basically obeyed, here we emphasize
that the possible mixed occupations in the $B/D$ site may greatly
influence the maximum $T_c$. In this regard, the simpler 1111 system
has the advantage avoiding such mutual occupations, and therefore it
exhibits the highest $T_c$ among FeSCs.

\begin{table}
\caption{\label{tab:table5}Four groups of "extended" iron-based
superconductors containing relatively thick intermediate layers.}
\begin{ruledtabular}
\begin{tabular}{lcr}
Structure/system&Chemical formula& $T_{c,max}$ (K)\\
\hline
 &$A_{n+1}B_{n}D_{2}X_{2}Z_{3n-1}$  &\\
32225 ($n$=2) & Sr$_3$(Sc,Ti)$_2$Fe$_2$As$_2$O$_5$ &20\cite{32225-chengf}\\
32225 ($n$=2)& Ca$_3$Al$_2$Fe$_2$As$_2$O$_5$ & 30.2\cite{32225-shirage}\\
43228 ($n$=3)& Ca$_4$(Sc,Ti)$_3$Fe$_2$As$_2$O$_8$ & 47\cite{42238-ogino}\\
5422\underline{11} ($n$=4)& Ca$_5$(Sc,Ti)$_4$Fe$_2$As$_2$O$_{11}$ & 46 \cite{thicker-ogino}\\
6522\underline{14} ($n$=5)& Ca$_6$(Sc,Ti)$_5$Fe$_2$As$_2$O$_{14}$ & 42 \cite{thicker-ogino}\\
\hline
 &$A_{n+2}B_{n}D_{2}X_{2}Z_{3n}$ &  \\
21113 ($n$=2) & Sr$_2$ScFePO$_3$ & 17\cite{21113P}\\
21113 ($n$=2) & Sr$_2$VFeAsO$_3$ & 37.2\cite{21113-whh}\\
32116 ($n$=4) & Ca$_3$(Al,Ti)$_2$FeAsO$_6$ & 39\cite{53229}\\
\hline
10-3-8&Ca$_{10}$Pt$_3$As$_8$(Fe$_2$As$_2$)$_5$& 11\cite{Pt-cava}\\
10-4-8&Ca$_{10}$Pt$_4$As$_8$(Fe$_2$As$_2$)$_5$& 26\cite{Pt-cava}\\
\hline
22241 & Ba$_2$Ti$_2$Fe$_2$As$_4$O & 21\cite{syl}\\
\end{tabular}
\end{ruledtabular}
\end{table}

\subsection{Superconductivity induced by chemical doping}

Most FeSCs are realized by a certain chemical doping, although there
are few exceptions (e.g., $\beta$-FeSe and LiFeAs). The parent
compounds are generally an antiferromagnetic spin-density-wave (SDW)
bad metal. By chemical doping, the SDW ordering is suppressed, and
then superconductivity emerges. For different systems and/or
different doping strategies, the details of the superconducting
phase diagram varies to some extent, but the generic tendency keeps
unchanged.\cite{rev-johnston,rev-stewart,rev-greene,rev-chucw} It is
widely believed that such generic phase diagram holds even for other
exotic superconductors including high $T_c$ cuprates, high $T_c$
pnictides and chalcogenides, heavy Fermion superconductors and
organic superconductors.

However, the parent compounds of FeSCs seem to be most close to
superconducting ground state, because various doping (being viewed
as a perturbation), even applying a moderate pressure, can induce
superconductivity. Table~\ref{tab:table6} summarizes the
representative chemical doping strategies that have been widely
employed. Several points can be drawn from the table. (1) Chemical
doping at \emph{any} crystallographic site is effective to induce
superconductivity. Of particularly surprising is that the doping
(even up to 50\%) within Fe$_2$As$_2$ layers does not destroy
superconductivity. (2) Superconductivity is able to be introduced by
either electron doping, hole doping, or non-charge-carrier
(so-called isoelectronic) doping. (3) Superconductivity can be
observed without doping. In this category, we believe that either
internal charge transfer (self-doping\cite{21113-cao,jh}) or
internal pressure is at work. (4) Overall speaking, electron doping
is easier to obtain superconductivity. For detailed information to
the chemical doping study, a featured review (in Chinese) can be
referred.\cite{rev-wc}

\begin{table}
\caption{\label{tab:table6}Classification of chemical doping that
induces Fe-based superconductivity in a $A-B-D-X-Z$ (see text for the notations) system.}
\begin{ruledtabular}
\begin{tabular}{lccr}
Site/Type&Electron&Hole&Isoelectronic\\
\hline
$A$& Th$^{4+}$/$R^{3+}$\cite{wc}& $A^{+}$/$A_{E}^{2+}$\cite{rotter}&Not successful\\
$B$& Ti$^{4+}$/Sc$^{3+}$ \cite{32225-chengf}&Not successful&Not successful\\
$D$& Co/Fe\cite{1111Co1,1111Co2}& Not successful& Ru/Fe\cite{122Ru}\\
 $X$& Not successful& Not successful& P/As\cite{js}\\
$Z$& F$^{-}$, H$^{-}$/O$^{2-}$\cite{hosono}& Not successful&Not successful\\
$Z$&O vacancy \cite{hosono}& Not successful&Not successful\\
\end{tabular}
\end{ruledtabular}
\end{table}

Table~\ref{tab:table6} also indicates that, in some cases, the
chemical doping is not successful to generate superconductivity. One
of the reasons is that the doping limit (or the substitution
solubility) is too small to turn superconductivity on. The other
reason for the absence of superconductivity is that, the doping
fails to supply sufficient perturbations to push the system to
superconducting ground state, or the dopant itself kills the
potential superconductivity. For example, the isoelectronic doping
at $A$ site may also supply chemical pressure (CP), but it is not
successful to induce superconductivity. We explain this phenomenon
as follows. In a multielement system, the CP may be "inhomogeneous",
as was demonstrated by a systematic study on
($R$,Pr)Ba$_2$Cu$_3$O$_y$-related system.\cite{Pr123a,Pr123b,Pr123c}
In iron pnictides, the CP by $A$-site doping does not supply enough
internal pressure onto the Fe$_2$$X_2$ layers (although the lattice
constants $a$ and $c$ both decrease). That is the possible reason
why the CP by the doping away from the Fe$_2$$X_2$ layers cannot
produce superconductivity.

\section{Structural design}

\subsection{Principles of structural design}

Since all the known FeSCs contain Fe$_2$$X_2$ layers, new FeSCs can
be explored by a rational structural design. The structural building
is something like a "lattice stacking engineering" (LSE).
Fortunately, the LSE for FeSCs is relatively simple because it is
confined to the stacking along the $c$ axis in order to be
compatible with the infinite  Fe$_2$As$_2$ layers. In practice, it
is our aim to find a distinct block layers between the Fe$_2$$X_2$
layers, and the designed object must be ultimately synthesized.

Here we propose some points that should be taken into considerations
for a structural design:

1) \emph{Laws of crystal chemistry.} In general, the law of crystal
chemistry\cite{Pauling} should be referred as a guidance. For
example, the radius ratio rule is particularly effective for an
ionic bonding.

2) \emph{Lattice Match.} Perhaps the first condition for a
successful LSE is that the building block should basically match the
essential Fe$_2$$X_2$ lattice. In this respect, Fe$_2$$X_2$ layers
have very good compatibility because the $a$ axis varies from
$\sim$3.75 {\AA} to $\sim$4.10 {\AA}. Therefore, many
crystallographic layers like CaF$_2$-type, CsCl-type, NaCl-type, and
perovskite-like blocks can be employed as candidates for the
building unit.

3) \emph{Charge Balance.} The next important issue to be considered
is charge balance. Empirically, the most stable valence state of Fe
in pnictides or chalcogenides is 2+. This means that the
Fe$_2$$Pn_2$ ($Pn$ stands for a pnictogen) unit carries charges
+2$e$ such that the intermediate block must bring approximate $-2e$
to neutralize the system. This is the case that most FeSCs obey. The
non-charged Fe$_2$$Ch_2$ ($Ch$ denotes a chalcogen) layers in the
chalcogenides makes it not so compatible for an ionic interlayer
bonding. For the collapsed 122 block, there is interlayer covalent
bonding, thus the change of the valence state of $X$ should be
considered.

4) \emph{Rule of HSAB.} The HSAB concept can be very useful in
designing a multielement compounds. With the classification of the
different type of elements in the HSAB scheme, one may easily find a
suitable element for a specific site.

5) \emph{Self stability.} Your designed compound must be at least
apparently stable by itself. Mixed occupations should be avoid as
far as possible. Make sure that "internal redox reaction" will not
take place in the chemical formula designed.

\subsection{Examples of structural design}

In this Subsection, we present some examples of structural design
for FeSCs. One should recognize that the proposed structures are not
successful unless they are ultimately synthesized. On the other
hand, it is very difficult to rule out the possiblity of realization
of a certain structure, because there are lots of possible
combinations of the constituent elements and, the synthetic methods
and conditions are not limited. We wish the readers would find that
some of the examples here could be useful and illuminating.

\begin{figure*}
\includegraphics[width=14cm]{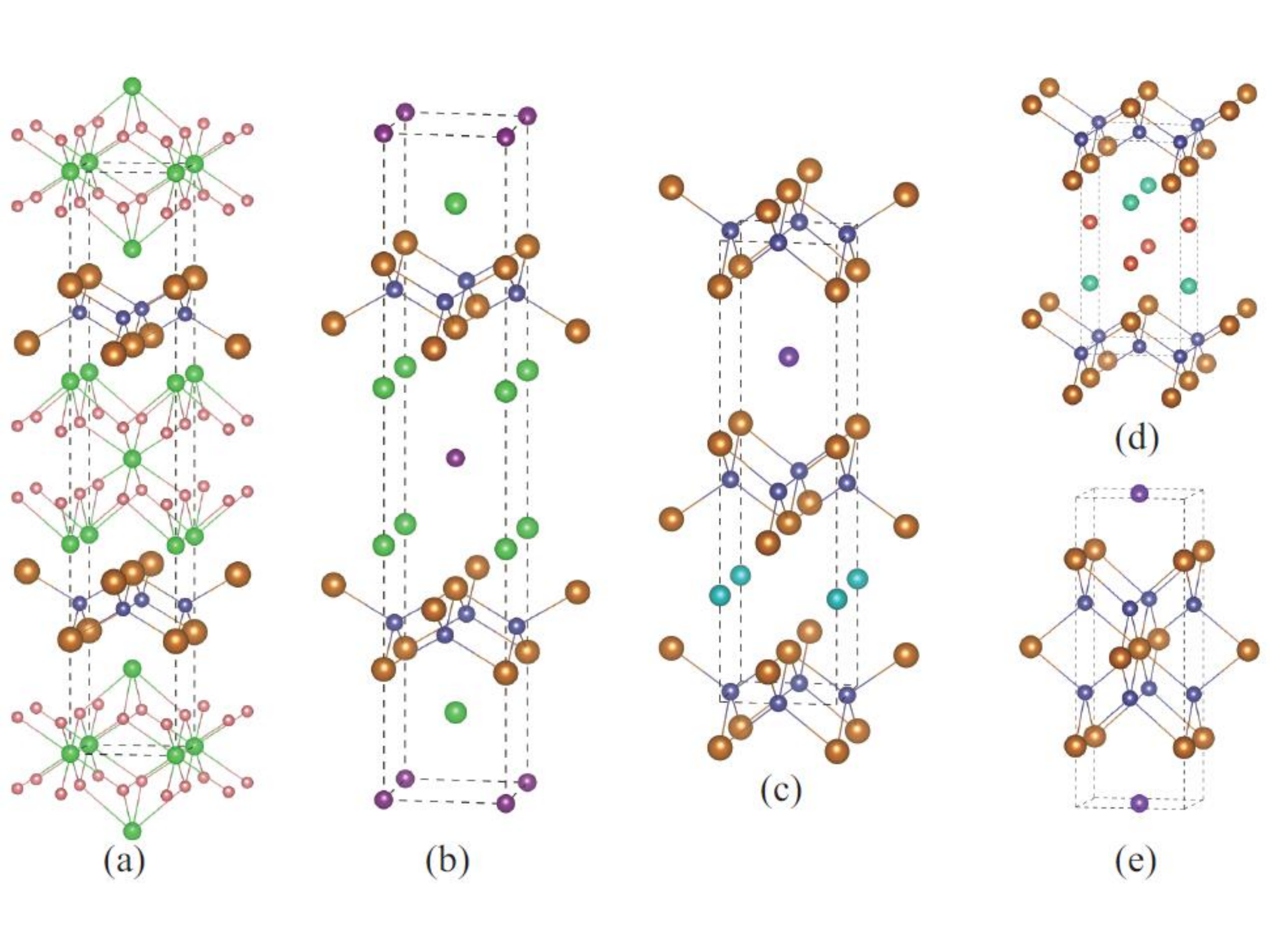}
\caption {\label{fig8} Examples of structural design for FeSCs. (a)
$A_{3}$Fe$_{2}X_{2}Z_{4}$; (b) $A_{2}$Fe$_{2}X_{2}X'$; (c)
KLaFe$_4$As$_4$; (d) $A$FeAsCl; (e) $A$Fe$_4$As$_3$.}
\end{figure*}

Fig.~\ref{fig8} displays some simple structures. Structure (a) is
derived by adding another fluorite-type layer on the 1111 phase. The
general formula is $A_{3}$Fe$_{2}X_{2}Z_{4}$. Such thick
fluorite-type layers are very common in the cuprate
superconductors.\cite{raveau} In addition, there is example of the
similar building block in Eu$_3$F$_4$S$_2$.\cite{EuFS} So, this
structure could be synthesized if one chose the correct compositions
and with suitable synthetic techniques (perhaps high-pressure method
should be employed).

By adding a CsCl-type $AX'$ layers into the 122 phase, the structure
of Fig.~\ref{fig8}(b) with a formula of $A_{2}$Fe$_{2}X_{2}X'$ can
be obtained. One should pay attention to the charge valence for this
structure. Fig.~\ref{fig8}(c) shows an example that the $A$-site
ions are ordered because of the difference in valence states or
ionic sizes. It is actually a superstructure (along the $c$ axis) of
122 phase.  A possible candidate is KLaFe$_4$As$_4$. Although our
preliminary attempt failed, it could be successful by a
low-temperature synthesis which tends to favor an ordered phase.

Up to present, the record $T_c$ of FeSCs appears in 1111 system. So,
a 1111-like structure would be promising for an elevated $T_c$. If
the $Z$-site ion of the 1111 phase is too large to adopt the
fluorite configuration, a distorted 1111 structure would form. This
case is shown in Fig.~\ref{fig8}(d). The intermediate layer is
analogous to the Van de Waals gap of PbFCl. So, $A$FeAsCl ($A$=Ca,
Cd, Pb) could be candidates of this type structure.

It seems to be true that the Fe$_2$$X_2$-based materials have a
$T_{c}$ limit of $\sim$60 K. To explore opportunities of higher
$T_{c}$ in FeSCs, one may consider modifying the Fe$_2$$X_2$ layers.
Fig.~\ref{fig8}(e) shows an example of this attempt. It contains a
thick superconductively-active Fe$_4$As$_3$ layers. If such a
material came out, higher $T_c$ could be expectable. It is noted
here that the same structure already exists (e.g.,
KCu$_4$S$_3$\cite{KCuS}).

\begin{figure*}
\includegraphics[width=14cm]{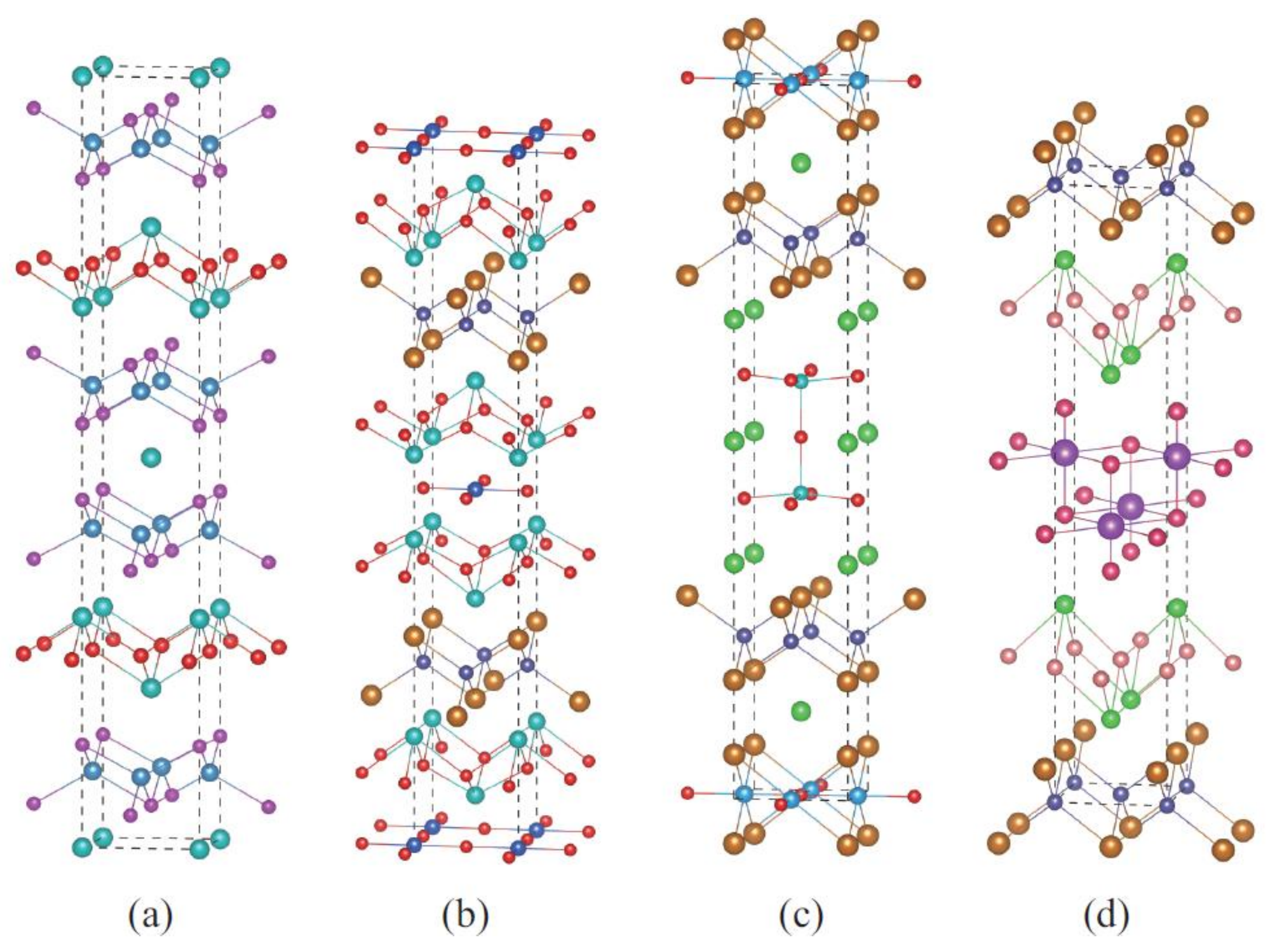}
\caption {\label{fig9} More examples of structural design for FeSCs.
(a) $A_{3}$Fe$_{4}X_{4}Z_{2}$; (b) $A_{4}B$Fe$_{2}X_{2}Z_{6}$; (c)
$A_{5}B_{4}$Fe$_{4}X_{4}Z_{7}$; (d) (La,Sr)$_2$InFeSe$_3$O$_2$.}
\end{figure*}

The structures shown in Fig.~\ref{fig9} are constructed by more
building blocks. The very left structure is an intergrowth of 122
and 1111. The resultant chemical formula is
$A_{3}$Fe$_{4}X_{4}Z_{2}$. A nickel-based material crystallizes in
the same structure, which shows superconductivity at 2.2
K.\cite{3442-cava}

In Fig.~\ref{fig9}(b), a thick $A_{4}BZ_{6}$ block layer, which
consists of $A_{2}BZ_{4}$ and $A_{2}Z_{2}$ layers, is sandwiched
into the Fe$_{2}X_2$ layers. The general chemical formula is
$A_{4}B$Fe$_{2}X_{2}Z_{6}$. Fig.~\ref{fig9}(c) shows a composite
structure containing Fe$_2$$X_2$, perovskite-like $A_{3}B_{2}Z_{5}$
and anti-perovskite $B_{2}X_{2}Z$ layers. The general chemical
formula is $A_{5}B_{4}$Fe$_{4}X_{4}Z_{7}$.

The last structure we recommend contains a building block
Bi$_{2}$S$_{4}$ which was recently found as a new
superconductively-active layers.\cite{BiS} In this proposed
structure, a fluorite-type $A_{2}Z_{2}$ layers are present. It can
also be viewed as an intergrowth of LaBiS$_2$O and 1111 structure. A possible
candidate is (La,Sr)$_2$InFeSe$_3$O$_2$, after considering the
lattice match and charge balance.

To be honest, most designed objects are not likely to be
successful. Thus the evaluation of the stability of a
designed material is very crucial. In this respect,
first-principles calculations would help under certain
circumstances. We hope to develop this kind of skills soon, in order to make the
structural design more rational.

\section{Summary and outlook}

In summary, the crystal chemistry of FeSCs bears some common
characteristic such that a structural design for new FeSCs is
possible. All the known FeSCs possess anti-fluorite-type Fe$_2$$X_2$
layers, which are believed to be responsible for superconductivity.
Other elements in a certain FeSC help to form a three-dimensional
lattice, and also assist to tune the superconductivity to an optimal
level.

There have been some dozens of FeSCs discovered so far, classified
into nine groups as listed in Table~\ref{tab:table1} and
Table~\ref{tab:table5}. The superconducting transition temperature
$T_c$ scatters from $\sim$10 K to $\sim$55 K, somewhat depending on
the structural parameters such as the bond angle of $X-$Fe$-X$ and
the height of $X$ with respect to the Fe-plane. In fact, the maximum
$T_c$ also depends on the type of doping. In general, the doping on
the Fe$_2$$X_2$ layers (especially at the Fe site) inevitably brings
disorder effect, which could suppress $T_c$ to some extent.

By employing the HSAB concept, it is more clear that a specific
crystallographic site in a FeSC belongs to a certain group of
elements. Based on this point, we have summarized the members of
iron-containing compounds for the major type of materials
(Table~\ref{tab:table2} to Table~\ref{tab:table4}). We have also
summarized different kinds of chemical doping that may induce
superconductivity (Table~\ref{tab:table6}). These knowledges should
be useful for selecting a chemical dopant and designing a new
structure.

It has been over five years since the breakthrough of FeSCs in the
Spring of 2008. The number of new FeSCs discovered seems to decay
rapidly with time. So, maybe we need a rational route in combination
with a systematic study. We have proposed nine structures here. It
is our hope that they could be synthesized at some time in the near
future.


\begin{thebibliography}{00}
\bibitem{onnes}Onnes H K 1911 \emph{Commun. Phys. Lab.} \textbf{12} 120

\bibitem{bednorz}Bednorz J G and Muller K A 1986 \emph{Z. Phys. B} \textbf{64} 189

\bibitem{chu}Wu M K, Ashburn Jr, Torng C J, Hor P H, Meng R L, Gao L, Huang Z J, Wang Y Q and Chu C W 1987 \emph{Phys. Rev. Lett.} \textbf{58} 908
\bibitem{hosono}Kamihara Y, Watanabe T, Hirano M and Hosono H 2008 \emph{J. Am. Chem. Soc.} \textbf{130} 3296
\bibitem{cxh}Chen X H, Wu T, Wu G, Liu R H, Chen H and Fang D F 2008 \emph{Nature} \textbf{453} 761

\bibitem{bcs}Bardeen J, Cooper L N and Schrieffer J R 1957 \emph{Phys. Rev.} \textbf{108} 1175

\bibitem{raveau}Raveau B, Michel C, Hervieu M, Groult D 1991 \emph{Crystal chemistry of high Tc superconducting copper oxides} (Springer-Verlag)

\bibitem{hosono2006}Kamihara Y, Hiramatsu H, Hirano M, Kawamura R, Yanagi H, Kamiya T and Hosono H 2006 \emph{J. Am. Chem. Soc.} \textbf{128} 10012
\bibitem{zzx}Ren Z A, Lu W, Yang J, Yi W, Shen X L, Li Z, Che G C, Dong X L, Sun L L, Zhou F and Zhao Z X 2008 \emph{Chin.
Phys. Lett.} \textbf{25} 2215
\bibitem{wc}Wang C, Li L, Chi S, Zhu Z, Ren Z, Li Y, Wang Y, Lin X, Luo Y, Jiang S, Xu X, Cao G and Xu Z 2008 \emph{EPL} \textbf{83} 67006
\bibitem{rotter}Rotter M, Tegel M and Johrendt D 2008 \emph{Phys. Rev. Lett.} \textbf{101} 107006
\bibitem{jcq}Wang X C, Liu Q Q, Lv Y X, Gao W B, Yang L X, Yu R C, Li F Y and Jin C Q 2008 \emph{Solid State Commun.} \textbf{148} 538
\bibitem{wumk}Hsu F C, Luo J Y, Yeh K W, Chen T K, Huang T W, Wu P M, Lee Y C, Huang Y L, Chu Y Y, Yan D C and Wu M
K 2008 \emph{Proc. Natl. Acad. Sci. USA} \textbf{105} 14262

\bibitem{rev-johnston}Johnston D C 2010 \emph{Adv. Phys.} \textbf{59} 803
\bibitem{rev-stewart}Stewart G R 2011 \emph{Rev. Mod. Phys.} \textbf{83} 1589
\bibitem{rev-hosono}Ishida K, Nakai Y and Hosono H 2009 \emph{J. Phys.
Soc. Jpn.} \textbf{78} 062001.
\bibitem{rev-zzx}Ren Z A and Zhao Z X 2009 \emph{Adv. Mater.} \textbf{21} 1
\bibitem{rev-whh}Wen H H 2008 \emph{Adv. Mater.} 20 3764
\bibitem{rev-wilson}Wilson J A 2010 \emph{J. Phys.: Condens. Matter} \textbf{22} 203201
\bibitem{rev-greene}Paglione J and Greene R L 2010 \emph{Nat. Phys.} \textbf{6} 645
\bibitem{rev-chucw}Chu C W 2009 \emph{Nat. Phys.} \textbf{5} 787
\bibitem{rev-mazin}Hirschfeld P J, Korshunov M M and Mazin I I 2011 \emph{Rep. Prog. Phys.}
\textbf{74} 124508
\bibitem{rev-ganguli}Ganguli A K, Prakash J and Thakur G S 2013 \emph{Chem. Soc. Rev.} \textbf{42} 569
\bibitem{chem-johrendt}Johrendt D, Hosono H, Hoffmann R D and P\"{o}ttgen R 2011
\emph{Z. Kristallogr.} \textbf{226} 435

\bibitem{pearson}Pearson R G 1963 \emph{J. Am. Chem. Soc.} \textbf{85} 3533
\bibitem{pottgen}P\"{o}ttgen R and Johrendt D 2008 \emph{Z. Naturforsch.} \textbf{63b} 1135
\bibitem{FeSe-PD}Schuster W, Milker H and Komarek K L 1979 \emph{Monatsh. Chem.} \textbf{110} 1153
\bibitem{X1}Zhang L J, Subedi A, Singh D J and Du M H \emph{Phys. Rev. B} \textbf{78} 174520
\bibitem{X2}Wang C, Ma Z F, Jiang S, Li Y K, Xu Z A and Cao G H 2010 Science China Physics, Mechanical \& Astronomy \textbf{53} 1225
\bibitem{X3}Muir S, Vielma J, Schneider G, Sleight A W and Subramanian M A 2012 \emph{J. Solid State Chem.} \textbf{185} 156

\bibitem{renza}Ren Z A, Yang J, Lu W, Yi W, Shen X L, Li Z C, Che G C, Dong X L, Sun L L, Zhou F and Zhao Z X
2008 \emph{EPL} \textbf{82} 57002
\bibitem{eisaki}Eisaki H, Iyo A, Kito H, Miyazawa K, Shirage P M, Matsuhata H, Kihou K, Lee C H, Takeshita N, Kumai R, Tomioka Y and Ito T 2008 \emph{J. Phys. Soc. Jpn.} \textbf{77} C36

\bibitem{zhaoj}Zhao J, Huang Q, Cruz C, Li S, Lynn J W, Chen Y, Green M A, Chen G F, Li G, Li Z, Luo J L, Wang N L and Dai P 2008 \emph{Nat. Mater.} \textbf{7} 953
\bibitem{lee}Lee C H, Iyo A, Eisaki H, Kito H, Fernandez-Diaz M T, Ito T, Kihou K, Matsuhata H, Braden M, Yamada K, 2008 \emph{J. Phys. Soc. Jpn.} \textbf{77} 083704
\bibitem{mizuguchi}Mizuguchi Y, Hara Y, Deguchi L, Tsuda S, Yamaguchi T, Takeda
K, Kotegawa H, Tou H and Takano Y 2010 \emph{Supercond. Sci.
Technol.} \textbf{23} 054013
\bibitem{kuroki}Kuroki K, Usui H, Onari S, Arita R and Aoki H 2009 \emph{Phys. Rev. B} \textbf{79} 224511

\bibitem{FeS}Taylor L A and Finger L W 1970 \emph{Carnegie Inst. Washington Geophys.
Lab. Annu. Rep.} \textbf{69} 318
\bibitem{FeSe}Hagg G and Kindstr\"{o}m A L 1933 \emph{Z. Phys. Chem. B} \textbf{22} 453
\bibitem{FeTe}Haraldsen H, Gr{\o}vold F and Vihovde J 1944 \emph{Tidsskr. Kjemi Bergves.} \textbf{4} 96
\bibitem{FeTe-crystal}Sales B C, Sefat A S, McGuire M A, Jin R Y, Mandrus D and Mozharivskyj Y 2009 \emph{Phys. Rev. B} \textbf{79} 094521
\bibitem{rev-takano}Mizuguchi Y and Takano Y 2010 \emph{J. Phys. Soc. Jpn.} \textbf{79} 102001
\bibitem{FeS1995}Lennie A R, Redfern S A T, Schofield P F and Vaughan D J 1995 \emph{Mineral. Mag.} \textbf{59} 677
\bibitem{Pauling}Pauling L 1960 \emph{The nature of the chemical bond, 3rd Ed.} (Cornell University Press)

\bibitem{11-fmh}Fang M H, Spinu L, Qian B, Pham H M, Liu T J, Vehstedt E K, Liu Y and Mao Z Q 2008 \emph{Phys. Rev. B} \textbf{78} 224503
\bibitem{FeSeTe}Yeh K W, Huang T W, Huang Y L, Chen T K, Hsu F C, Wu P M, Lee Y C, Chu Y Y, Chen C L, Luo J Y, Yan D C and Wu M K 2008 \emph{EPL} \textbf{84} 37002
\bibitem{FeTeS}Mizuguchi Y, Tomioka F, Tsuda S, Yamaguchi T and Takano Y 2009 \emph{Appl. Phys. Lett.} \textbf{94} 012503

\bibitem{11-cava}McQueen T M, Huang Q, Ksenofontov V, Felser C, Xu Q, Zandbergen H, Hor Y S, Allred J, Williams A J, Qu D, Checkelsky
J, Ong N P and Cava R J 2009 \emph{Phys. Rev. B} \textbf{79} 014522
\bibitem{11-HP1}Medvedev S, McQueen T M, Trojan I, Palasyuk T, Eremets M I, Cava R J, Naghavi S, Casper F, Ksenofontov V, Wortmann G and Felser C 2009 \emph{Nat. Mater.}
\textbf{8} 630
\bibitem{11-HP2}Margadonna S, Takabayashi Y, Ohishi Y, Mizuguchi Y, Takano Y, Kagayama T, Nakagawa T, Takata M and Prassides K 2009
\emph{Phys. Rev. B} \textbf{80} 064506

\bibitem{LiFeP-jcq}Deng Z, Wang X C, Liu Q Q, Zhang S J, Lv Y X, Zhu J L, Yu R C and Jin C Q 2009 \emph{EPL} \textbf{87} 37004

\bibitem{K122}Sasmal K, Lv B, Lorenz B, Guloy A M, Chen F, Xue Y Y and Chu C W 2008 \emph{Phys. Rev. Lett.} \textbf{101} 107007

\bibitem{122-hp}Torikachvili M S, Budko S L, Ni N and Canfield P C 2008 \emph{Phys. Rev. Lett.} \textbf{101} 057006
\bibitem{122Co}Sefat A S, Jin R, McGuire M A, Sales B C, Singh D J and Mandrus D 2008 \emph{Phys. Rev. Lett.} \textbf{101} 117004
\bibitem{js}Jiang S, Xing H, Xuan G, Wang C, Ren Z, Feng C, Dai J, Xu Z and Cao G 2009 \emph{J. Phys.: Condens. Matter} \textbf{21} 382203

\bibitem{cxl}Guo J, Jin S, Wang G, Zhu K, Zhou T, He M and Chen X 2010 \emph{Phys. Rev. B} \textbf{79} 180520(R)
\bibitem{122*-fmh}Fang M H, Wang H D, Dong C H, Li Z J, Feng C M, Chen J and Yuan H Q 2011 \emph{EPL} \textbf{94} 27009
\bibitem{sll}SunL, Chen X J, Guo J, Gao P, Huang Q Z, Wang H. Fang M, Chen X, Chen G, Wu Q, Zhang C, Gu D, Dong X, Wang L, Yang K, Li A, Dai X, Mao H K and Zhao Z 2012 \emph{Nature} \textbf{483} 67
\bibitem{1111-hp}Takahashi H, Igawa K, Arii K, Kamihara Y, Hirano M and Hosono H 2008 \emph{Nature} \textbf{453} 376

\bibitem{1111-whh}Wen H H, Mu G, Fang L, Yang H and Zhu X 2008 \emph{EPL} \textbf{82} 17009

\bibitem{1111Co1}Sefat A S, Huq A, McGuire M A, Jin R Y, Sales B C, Mandrus D, Cranswick L M D,
Stephens P W and Stone K H 2008 \emph{Phys. Rev. B} \textbf{78}
104505
\bibitem{1111Co2}Wang C, Li Y K, Zhu Z W, Jiang S, Lin X, Luo Y K, Chi S, Li L J, Ren
Z, He M, Chen H, Wang Y T, Tao Q, Cao G H and Xu Z A 2009
\emph{Phys. Rev. B} \textbf{79} 054521

\bibitem{1111P}Wang C, Jiang S, Tao Q, Ren Z, Li Y, Li L, Feng C,
Dai J, Cao G and Xu Z 2009 \emph{EPL} \textbf{86} 47002
\bibitem{CuFeSb-mzq}Qian B, Lee J, Hu J, Wang G C, Kumar P, Fang M H, Liu T J, Fobes D, Pham H, Spinu L, Wu X S, Green M,
Lee S H and Mao Z Q 2012 \emph{Phys. Rev. B} \textbf{85} 144427

\bibitem{111str}Pearson W B 1964 \emph{Can. J. Phys.} \textbf{42}, 519

\bibitem{NaFeAs-wnl}Chen G, Hu W, Luo J and Wang N 2009 \emph{Phys. Rev. Lett.} \textbf{102} 227004

\bibitem{MgFeGe}Welter R, Malaman B and Venturini G 1998 \emph{Solid State Comm.} \textbf{108} 933;
Hlukhyy V, Chumalo N, Zaremba V and F\"{a}sslera T F 2008 \emph{Z.
Anorg. Allg. Chem.} \textbf{634} 1249
\bibitem{MgFeGe-hosono}Liu X, Matsuishi S, Fujitsu S and Hosono H 2012 \emph{Phys. Rev.
B} \textbf{85} 104403

\bibitem{111-cxh}Wang A F, Xiang Z J, Ying J J, Yan Y J, Cheng P, Ye G J, Luo X G and Chen X H 2012 \emph{New J. Phys.} \textbf{14} 113043
\bibitem{111-hp}Zhang S J et al 2009 \emph{EPL} \textbf{88} 47008
\bibitem{111-pitcher}Pitcher M J, Parker D R, Adamson P, Herkelrath S J C, Boothroyd A T, Ibberson R M, Brunelli M and Clarke S J 2008 \emph{Chem. Commun.} 5918

\bibitem{122review}Just G and Paufler P 1996 \emph{J. Alloys Compd.} \textbf{232} 1

\bibitem{renz}Ren Z, Tao Q, Jiang S, Feng C M, Wang C, Dai J H, Cao G H and Xu Z A 2009 \emph{Phys. Rev. Lett.} \textbf{102} 137002
\bibitem{jiangs}Jiang S, Xing H, Xuan G F, Ren Z, Wang C, Xu Z A and Cao G H 2009 \emph{Phys. Rev. B} \textbf{80} 184514
\bibitem{jwh}Jiao W H, Tao Q, Bao J K, Sun Y L, Feng C M, Xu Z A, Nowik I, Felner I and Cao G H 2011 \emph{EPL} \textbf{95} 67007
\bibitem{cao}Cao G H, Xu S G, Ren Z, Jiang S, Feng C M and Xu Z A 2011 \emph{J. Phys. Condens. Matter} \textbf{23} 464204
\bibitem{CaLa122}Lv B, Deng L Z, Gooch M, Wei F Y, Sun Y Y, Meen J K, Xue Y Y, Lorenz B and Chu C W 2011 \emph{Proc. Natl. Acad. Sci. USA} \textbf{108} 15705

\bibitem{122collapse}Hoffmann R and Zheng C 1985 \emph{J. Phys. Chem.} \textbf{89} 4175
\bibitem{CaBeGe}Eisenmann B, May N, M\"{u}ller W and Sch\"{a}fer H Z 1972 Z.
Naturforsch. B \textbf{27} 1155

\bibitem{SrPtAs}Kudo K, Nishikubo Y and Nohara M 2010 \emph{J. Phys.
Soc. Jpn.} \textbf{79} 123710

\bibitem{rev-whh2012}Wen H H 2012 \emph{Rep. Prog. Phys.} \textbf{75} 112501

\bibitem{1111-quebe}Quebe P, Terb\"{u}chte L J and Jeitschko W 2000 \emph{J. Alloys Compd.} \textbf{302} 70
\bibitem{Np1111}Klimczuk T, Walker H C, Springell R, Shick A B, Hill A H,Gaczy\'{n}ski P, Gofryk K, Kimber S A J
Ritter C, Colineau E, Griveau J C, Bou\"{e}xi\`{e}re D, Eloirdi R,
Cava R J and Caciuffo R 2012 \emph{Phys. Rev. B} \textbf{85} 174506

\bibitem{1111H-hosono}Hosono H and Matsuishi S 2013 arXiv: 1304.7900

\bibitem{21222-zhu}Zhu W J and Hor P H 1997 \emph{J. Solid State Chem.} \textbf{130} 319;
Zhu W J, Hor P H, Jacobson A J, Crisci G, Albright T A, Wang S H and
Vogt T 1997 \emph{J. Am. Chem. Soc.} \textbf{119} 12398
\bibitem{32225-zhu}Zhu W J, and Hor P H 1997 \emph{J. Solid State Chem.} \textbf{134} 128
\bibitem{43228-zhu}Zhu W J and Hor P H 2000 \emph{J. Solid State Chem.} \textbf{153} 26

\bibitem{32225-kishio}Otzschi K, Ogino H, Shimoyama J and Kishio K 1999 \emph{J. Low Temp. Phys.} \textbf{117} 729

\bibitem{43228-hyett}Hyett G, Barrier N, Clarke S J and Hadermann J 2007 \emph{J. Am. Chem. Soc.} \textbf{129} 11192

\bibitem{32225-whh}Zhu X, Han F, Mu G, Zeng B, Cheng P, Shen B and Wen H H 2009 \emph{Phys. Rev. B} \textbf{79} 024516

\bibitem{32225-chengf}Chen G F, Xia T L, Yang H X, Li J Q, Zheng P, Luo J L and Wang N L 2009 \emph{Supercond. Sci.
Technol.} \textbf{22} 072001

\bibitem{32225-shirage}Shirage P M, Kihou K, Lee C H, Kito H, Eisaki H and Iyo A 2011 \emph{J. Am. Chem. Soc.} \textbf{133} 9630
\bibitem{42238-ogino}Ogino H, Shimizu Y, Ushiyama K, Kawaguchi N, Kishio K and Shimoyama J 2010 \emph{Appl. Phys.
Express} \textbf{3} 063103
\bibitem{thicker-ogino}Ogino H, Sato S, Kishio K, Shimoyama J, Tohei T and Ikuhara Y 2010 \emph{Appl. Phys. Lett.} \textbf{97} 072506
\bibitem{master-thesis}Yu J 2010 First-principles electronic structure Calculations for Ba$_2$CuO$_2$Fe$_2$As$_2$ (MS Thesis) (Hangzhou: Zhejiang
University) (in Chinese)
\bibitem{21222-nath}Nath R, Garlea V O, Goldman A I and Johnston D C 2010 \emph{Phys. Rev. B} \textbf{81} 224513
\bibitem{21222Cr}Eguchi N, Ishikawa F, Kodama M, Wakabayashi T,
Nakayama A, Ohmura A and Yamada Y 2013 \emph{J. Phys. Soc. Jpn.}
\textbf{82} 045002.
\bibitem{53229}Ogino H, Machida K, Yamamoto A, Kishio K, Shimoyama J. Tohei T and Ikuhara Y 2010 \emph{Supercond. Sci. Technol.} \textbf{23} 115005

\bibitem{RP}Ruddlesden S N and Popper P 1958 \emph{Acta Crystallogr.} \textbf{11} 54

\bibitem{21113-zhu}Zhu W J and Hor P H 1997 \emph{Inorg. Chem.} \textbf{36} 3576
\bibitem{21113P}Ogino H, Matsumura Y, Katsura Y, Ushiyama K, Horii
S, Kishio K and Shimoyama J 2009 \emph{Supercond. Sci. Technol.}
\textbf{22} 075008
\bibitem{21113-whh}Zhu X Y, Han F, Mu G, Cheng P, Shen B, Zeng B and Wen H H 2009 \emph{Phys. Rev. B} \textbf{79} 220512(R)
\bibitem{21113-cao}Cao G, Ma Z, Wang C, Sun Y, Bao J, Jiang S, Luo Y, Feng C, Zhou Y, Xie Z, Hu F, Wei S, Nowik I, Felner I,
Zhang L, Xu Z and Zhang F 2010 \emph{Phys. Rev. B} \textbf{82}
104518

\bibitem{21113Sc}Ogino H, Katsura Y, Horii S, Kishio K and Shimoyama J 2009
\emph{Supercond. Sci. Technol.} \textbf{22} 085001
\bibitem{21113Cr}Tegel M, Hummel F, Su Y, Chatterji T, Brunelli M and Johrendt D
2010 \emph{EPL} \textbf{89} 37006

%\bibitem{438Pt}cava
\bibitem{Pt-cava}Ni N, Allred J M, Chan B C and Cava R J 2011 \emph{Proc. Natl. Acad. Sci. USA} \textbf{108} E1019

\bibitem{ozawa-review}Ozawa T C and Kauzlarich S M 2008 \emph{Sci. Technol. Adv. Mater.} \textbf{9} 033003
\bibitem{1221-cxh}Wang X F, Yan Y J, Ying J J, Li J Q, Zhang M, Xu N and Chen X H 2010 \emph{J. Phys.: Condens. Matter} \textbf{22} 075702

\bibitem{syl}Sun Y L, Jiang H, Zhai H F, Bao J K, Jiao W H, Tao Q, Shen C Y, Zeng Y W, Xu Z A and Cao G H 2012 \emph{J. Am. Chem. Soc.} \textbf{134} 12893
\bibitem{jh}Jiang H, Sun Y L, Dai J H, Cao G H and Cao C 2012 arXiv 1207.6705

\bibitem{122Ru}Sharma S, Bharathi A, Chandra S et al 2010 \emph{Phys.
Rev. B} \textbf{81} 174512

\bibitem{rev-wc}Wang C, Cao G H and Xu Z A 2010 \emph{Prog. in Phys. (in Chinese)} \textbf{30} 307
\bibitem{Pr123a}Cao G H, Qian Y T, Chen Z Y, Li X J, Wu H K and Zhang Y
H 1994 \emph{Phys. Lett. A} \textbf{196} 263
\bibitem{Pr123b}Cao G H, Qian Y T, Li X J, Chen Z Y, Wang C Y, Ruan K Q, Qiu Y M, Cao L Z,
Ge Y A and Zhang Y H 1995 \emph{J. Phys.: Condens. Matter}
\textbf{7} L287
\bibitem{Pr123c}Song H Q, Cao G H, Xu Z A, Ren Y H and Zhang Q R 1997
\emph{Zeitschrift Physik B} \textbf{103} 29
\bibitem{EuFS}Grossholz H, Hartenbach I, Kotzyba G, P\"{o}ttgen R, Trill H, Mosel
B D and Schleid T 2009 \emph{J. Solid State Chem.} \textbf{182} 3071

\bibitem{KCuS}R\"{u}dorff W, Schwarz H G and Walter M 1952 \emph{Z. Anorg. Allg.
Chem.} \textbf{269} 141
\bibitem{3442-cava}Klimczuk T, McQueen T M, Williams A J, Huang Q, Ronning F, Bauer E D, Thompson J D, Green M A and Cava R J 2009 \emph{Phys. Rev. B} \textbf{79} 012505

\bibitem{BiS}Mizuguchi Y, Fujihisa H, Gotoh Y, Suzuki K, Usui H,
Kuroki K, Demura S, Takano Y, Izawa H and Miura O 2012 \emph{Phys.
Rev. B} \textbf{86} 220510(R)


\end{thebibliography}
\end{document}